\documentclass[twocolumn]{autart}    
\usepackage{amsmath}
\usepackage{amsfonts} 
\usepackage{amssymb}
\usepackage{graphicx}
\usepackage{cite}
\usepackage{color,xcolor}

\usepackage{epsf,psfrag,verbatim,epstopdf,epsfig,url,subfig}

\newcommand{\argmin}{\mathop{\mathrm{argmin}}}

\def\scr#1{{\mathcal #1}}
\newcommand{\R}{\mathbb{R}}
\def\eq#1{\begin{equation}#1\end{equation}}
\def\rep#1{(\ref{#1})}
\newcommand{\bbb}{\mathbb}
\newtheorem{theorem}{Theorem}
\newtheorem{lemma}{Lemma}
\newtheorem{remark}{Remark}
\newtheorem{proposition}{Proposition}
\newtheorem{corollary}{Corollary}

\newtheorem{assumption}{Assumption}

\def\qed{ \rule{.08in}{.08in}}

\newcommand{\kai}[1]{\textcolor{black}{#1}}

\begin{document}

\begin{frontmatter}

\title{Distributed Learning of Average Belief \\Over Networks Using     Sequential Observations \thanksref{footnoteinfo}} 

\thanks[footnoteinfo]{This paper builds on some earlier results presented at the 2017 American Control Conference \cite{acc17}. Corresponding author: Kaiqing Zhang.}

\author{Kaiqing Zhang,
        Yang Liu,
        Ji Liu,
        Mingyan Liu,
        Tamer Ba\c{s}ar}
\thanks{K. Zhang and T. Ba\c{s}ar are with the Department of Electrical and Computer Engineering at   
University of Illinois at Urbana-Champaign (\texttt{\{kzhang66, basar1\}@illinois.edu}). 
Y. Liu is with the School of Engineering and Applied Sciences at Harvard University and the Department of Computer Science and Engineering at University of California, Santa Cruz (\texttt{yangliu@ucsc.edu}). 
J. Liu is with the Department of Electrical and Computer Engineering at Stony Brook University ({\tt ji.liu@stonybrook.edu}).
M. Liu is with the Electrical Engineering and Computer Science Department at University of Michigan (\texttt{mingyan@eecs.umich.edu}).
}

\begin{abstract}                          
This paper addresses the problem of distributed learning of average belief  with sequential observations, 
in which a network of $n>1$ agents aim to
reach a consensus on the average value of their beliefs, by exchanging information only with their neighbors. 
Each agent has sequentially arriving samples of its belief in an online manner.  The neighbor relationships among the $n$ agents are described by a  graph which is possibly time-varying, 
whose vertices correspond to agents and whose edges depict neighbor relationships.
Two distributed online algorithms are introduced for undirected and directed graphs, which  \kai{are both shown to converge to the average belief almost surely. {\color{black}Moreover, the sequences generated by both algorithms are shown to reach consensus with an $O({1}/{t})$  rate with high  probability,} where $t$ is the number of iterations.}   
For undirected graphs, the corresponding algorithm is modified for the case with quantized communication and limited precision of the division operation.
It is shown that the modified algorithm causes all $n$ agents to either reach a quantized consensus or enter a small neighborhood around the average of their beliefs. 
Numerical simulations are then provided to corroborate the theoretical results. 
\end{abstract}

\end{frontmatter}

\section{Introduction}
Considerable interest in developing algorithms for  distributed computation and decision making problems of all types has arisen over the past few decades, including consensus problems \cite{Morse2003},
multi-agent coverage problems \cite{Bullo2004}, power distribution system management  \cite{zhang18dynamica,zhang18dynamicb}, and  multi-robot formation control \cite{Francis2009}.
These problems have found applications in different fields, including sensor networks \cite{kar},
robotic teams \cite{Francis2009}, social networks \cite{Morse2012}, internet of things \cite{chen2018pieee}, and electric power grids \cite{zhang18distributed,zhang18dynamica}. 
For large-scale complex networks, 
distributed computation and control  are especially promising,  
thanks to their attractive features of 
fault tolerance and cost saving, and their ability to accommodate various physical constraints such as
 limitations on sensing, computation, and communication.


Among the distributed control and computation problems, the consensus problem 
\cite{Morse2003,Basar2007} 
is one of the  most basic and important task.
In a typical consensus process, the agents
in a given group all try to agree on some quantity by communicating
what they know only to their neighboring agents.
In particular, 
one important type of consensus process, called distributed averaging \cite{Boyd2004}, aims  to compute the average of the initial values of the quantity of interest to the agents. 
Existing work has  developed elegant solutions to such   conventional distributed averaging problems, such as  linear iterations \cite{metro2}; gossiping \cite{Boyd2006,pieee};
 push-sum \cite{kempe},  also known
as weighted gossip \cite{weighted}; ratio consensus \cite{ratio}; and double linear iterations \cite{acc12}.
%


In the present work, we extend the conventional distributed averaging problem setting to the case where each distributed agent has its local belief/measurement arriving sequentially. 
In the previous studies of distributed averaging,  each agent $i$ was assumed to hold $x_i(1)$ at initial time $t=1$, which corresponds to  the true belief, and the subsequent averaging and communication processes are carried out entirely over $x_i(1)$, $i\in\{1,2,\ldots,n\}$.  In contrast, we consider here the case where a series of local observations, denoted by $x_i(t)$, $t\in\{1,2,\ldots\}$, is available to each agent $i$, which are used to estimate the unknown true belief $\bar{x}_i$.
We refer to this setting as {\it distributed learning of average belief  using sequential observations}.

{\color{black}
Consider the motivating example of a sensor network, where each sensor needs to take a sequence of local measurements in order to obtain an accurate estimate of a local mean
(or belief as referred to earlier) due to environmental and instrumentation noises; at the same time we have the goal of
estimating the global/field mean through message exchange among sensors. It is clear that in this case there are multiple
averaging processes taking place  simultaneously, one globally among sensors, and a local one at each sensor. This is the
main difference between our problem and traditional distributed averaging which only focuses on the global averaging process by assuming that the local mean
is already available.
}

In general, the setting   considered here belongs to the  family of problems on distributed learning and control. 
We note that the distributed learning settings with noisy observations have  also been investigated in several   previous studies, e.g. \cite{huang,kar}.
The key differences between our work here and those studies are 
the following: (1) in our formulation, uncertainties are not modeled as coming  
 from external noise sources independent
of the sample observations, as the case in 
the previous study  \cite{huang}; 
(2) our algorithms are not developed based on a
distributed estimation framework as in \cite{kar},  where the  observability of
the system need to be assumed. 
Therefore, we note that there are multiple
averaging processes going on simultaneously in our  setting---a global one among the 
agents, and a local one at each agent. 
{\color{black} Similar ideas have been exploited in \cite{kar2013consensus+} and \cite{rahimian2016distributed} for two different problems, distributed inference and estimation, respectively.}
In the classical literature on
distributed averaging and consensus, however, only the  global averaging process is considered since  the local mean
is already available. 
It would be desirable to
embed the multiple underlying averaging processes here into the
same  updating procedure as in the classical distributed averaging process, without much modification.  
In particular, we should manage to integrate the new
measurements/samples that occur in an online fashion into   the classical distributed averaging process, which serves as  the goal of the present work. To this end,
we introduce two distributed online algorithms  and formally establish
their convergence and  convergence rates.

To implement the algorithms, the agents are required to  send, receive, as well as evaluate the running average of local beliefs with infinite precision. However, in a realistic network of agents,  messages with only
limited length are allowed to be transmitted among agents due to the capacity constraints of communication links. This is usually referred to as \emph{quantized communication} in distributed averaging, see previous  works on reaching   quantized consensus under such quantization \cite{Nedic2009b,carli2010gossip,chamie}.  Additionally, in the distributed belief averaging algorithms considered here, limited precision of belief averages  may occur due to the \emph{division operation}  in the local update at each agent. This is similar but more involved than the previous works on distributed averaging 
with integer/quantized  values at each agent \cite{Basar2007,cai2011quantized,bacsary2014fast}. We thus discuss the convergence of the proposed algorithms in the presence of these two quantization effects. \kai{We can show that  under certain conditions, the quantized update can converge to a small neighborhood of the actual average belief with bounded errors, even with such joint quantization effects.}



\kai{
The main contributions of this paper are threefold.
First, we propose a new setting of  distributed averaging  using sequential observations, and develop two easily-implementable  distributed algorithms for  undirected and directed graphs, respectively.
Second, we    establish almost sure convergence and polynomial convergence  rate with high probability for both algorithms.  
In addition, we investigate the effects of quantized communication and limited precision of the division operation on the algorithm for undirected graphs, 
and provide a convergence analysis under certain conditions.
}

This paper builds on some earlier results  presented in \cite{acc17}, but presents a more comprehensive
treatment of the problem. Specifically, the paper establishes the convergence rate for the algorithm over directed graphs, and characterizes the effects of two types of quantization on the algorithm over undirected graphs, which were not included in \cite{acc17}.

The rest of the paper is organized as follows.
The problem is formulated and stated in Section \ref{sec:pf}.
Two algorithms are presented to solve the problem  over  undirected and directed  graphs,  
 in  Section \ref{sec:static}, along with   results on their convergence rates. 
In Section \ref{sec:quan}, the convergence results for the algorithm under system quantization are provided, for both static and dynamic graphs. 
The analysis of the algorithms and  proofs of the main results are given in Section \ref{proof}. {\kai{The theoretical results are verified by numerical simulations in  Section \ref{sec:simu}, followed by  conclusions in Section \ref{sec:conclude}}.

\section{Problem Formulation}\label{sec:pf}

Consider a network consisting of $n>1$ agents, with the set of  agents  denoted by $\mathcal{N}=\{1,2,\cdots,n\}$. 
The neighbor relationships among the $n$ agents are described by a time-varying $n$-vertex  graph $\bbb{N}(t)=\left(\mathcal{N},\scr{E}(t)\right)$, called the {\em neighbor graph}, with $\scr{E}(t)$ denoting the set of edges at time $t$. Note that the graph $\bbb{N}(t)$ can be either undirected or directed. 
The $n$ vertices represent the $n$ agents and the edges indicate the neighbor relationships.
Specifically, if the communications among agents are bidirectional, the graph $\bbb{N}(t)$ is undirected, and agents $i$ and $j$ are neighbors at time $t$ if and only if
$(i,j)$ is an edge in $\bbb{N}(t)$. We use $\scr{N}_i(t)$ to denote the set of neighbors of agent $i$, i.e., the degree of vertex $i$ at time $t$. 
We also define $\scr{N}_i^+(t):=\mathcal N_i(t)\bigcup\{i\}$ as the set of neighbors including the agent $i$ itself.  Otherwise, if the communications are unidirectional, we say that an agent $j$ is an {\em out-neighbor} of agent $i$ at time $t$ if
agent $i$ can send information to agent $j$ at time $t$.
In this case, we also say that agent $i$ is an {\em in-neighbor} of agent $j$ at time $t$. 
Then, agent $j$ is a out-neighbor of agent $i$ (and thus agent $i$ is an in-neighbor of agent $j$) at time $t$
if and only if $(i,j)$ is a directed edge in $\bbb{N}(t)$. 
Thus, the neighbor graph $\bbb{N}(t)$ becomes directed, in which
the directions of  edges represent the directions of information flow.
By a slight   abuse of notation, we use $\scr N_i(t)$  to denote the set of out-neighbors of agent $i$ at time $t$, and also let $\scr{N}_i^+(t)=\scr N_i(t)\bigcup\{i\}$.
We also  use $o_i(t)$ to denote the number of out-neighbors   of agent $i$ at time $t$, or equivalently,
the out-degree of vertex $i$ in $\bbb{N}(t)$.
 
   


We assume that time is discrete in that $t$ takes values in $\{1,2,\ldots\}$.
Each agent $i$ receives a real-valued scalar\footnote{
The results in this paper can be straightforwardly extended to the vector-valued case.}
$x_i(t)$ at each time $t$.
We assume that the samples $\{x_i(t)\}_{t=1}^{\infty}$ form an independent and identically distributed (i.i.d.)  process, and
the sequence is generated according to a random variable $X_i$ with distribution $f_{X_i}(\cdot)$. For simplicity, we assume that the support set for $X_i$ is bounded, i.e., there exists a constant $K$ such that
$|X_i(\omega)| \leq K$ for all $i\in\scr{N}$ and $\omega$,
where $\omega$ indicates an arbitrary sample realization. Note that the agents' observations do not need to be identical, i.e., the $f_{X_i}(\cdot)$, $i\in\scr N$, do not need to be structurally the same.
We use $\bar{x}_i$ to denote the expectation of agent $i$'s local observations, i.e.,
$$\bar{x}_i=E[X_i]~,$$
and call $\bar x_i$ the {\em local belief} of agent $i$.
{\color{black}
An application context for this problem in sensor networks was described  in the introduction as a motivating example.  
}

At each time step $t$, each agent $i$ can exchange information only with its current neighbors $j \in \mathcal N_i(t)$.
Thus, only local information is available to each agent, i.e., each agent $i$, only knows
its own samples, the information received from its current neighbors in $\mathcal N_i(t)$, and nothing more, while the global connectivity patterns remain unknown to any agent in the network.

Suppose that each agent has control over a real-valued variable $y_i(t)$
which it can update from time to time.
Let
\eq{\bar{x}=\frac{1}{n}\sum_{i=1}^n \bar{x}_i~,\label{bar}}
 represent the {\em average belief} within the network.
Our goal is to devise distributed  algorithms for each agent $i$,   over either undirected or directed graphs, which ensures that
$$
\lim_{t\rightarrow \infty} y_i(t) = \bar{x}~, \;\;\;\;\; \forall i~, 
$$
\kai{almost surely (a.s.).}
Moreover, we would like  to characterize the convergence properties  
of such  algorithms 
\kai{with high probability (w.h.p.)\footnote{
Throughout the paper, when we say  with high probability, we mean  that the probability goes to $1$ when $t$ goes to infinity. 
\kai{
Note that this   is akin to the standard concept of ``convergence in probability'' \cite{durrett2010probability}. 
In particular, our results of   \emph{reaching consensus with ${O}(1/t)$  rate w.h.p.}  not only imply \emph{reaching consensus  in probability}, but also provide the \emph{convergence rate} of the sequence as $t\to \infty$. }
}}}, and their extensions to the case with   quantization effects of the system, as we will elaborate next.


\section{Distributed Learning  Algorithms}\label{sec:static}

In this section, we  introduce  two algorithms for distributed learning of average belief using sequential observations over time-varying graphs. 
We establish  \kai{both almost sure convergence and    ${O}(1/t)$ convergence rate w.h.p. for the algorithms.}  

\subsection{Algorithms}
We first consider the case where $\bbb{N}(t)$ is a time-varying undirected graph.
At initial time $t=1$, each agent $i$ sets  $y_i(1)=x_i(t)$, where we note that  $x_i(t)$ is an i.i.d. sample of the local belief ${X}_i$.
For each time $t>1$, each agent $i$ first broadcasts its current $y_i(t)$ to all its current neighbors
in $\scr{N}_i(t)$.
At the same time, each agent $i$ receives $y_j(t)$ from all its neighbors $j\in\scr{N}_i(t)$.
Then, each agent $i$ updates its variable by setting
\begin{eqnarray}
y_i(t+1) &=& w_{ii}(t)y_i(t) + \sum_{j\in\scr{N}_i(t)} w_{ij}(t)y_j(t) \nonumber \\
&&+ z_i(t+1) - z_i(t)~, \label{update}
\end{eqnarray}
where
\eq{z_i(t) = \frac{1}{t}\sum_{\tau=1}^tx_i(\tau) \label{equ:ave_z}}
is the \emph{running average} of the data, and $w_{ij}(t)$ is the real-valued weight  from the  matrix $W(t)=[w_{ij}(t)]\in\mathbb{R}^{n\times n}$. 
The update rule \eqref{update} can be written in a more compact form as 
\eq{y(t+1) = W(t)y(t)+z(t+1)-z(t)~, \label{state}}
where $y(t), z(t)\in \mathbb{R}^n$ are the column vectors obtained by stacking up $y_i(t)$s and $z_i(t)$s, respectively.

We make the following standard assumptions on $W(t)$ hereafter unless stated otherwise.

	\begin{assumption}\label{assum:Doubly_Sto}
	The weight matrix $W(t)$ has the following properties at any time $t$.
	
	A.1) $W(t)$ is a symmetric stochastic matrix\footnote{A square nonnegative matrix is called stochastic if its row sums all equal~$1$, and called doubly stochastic if its row sums and column sums all equal $1$. Thus, $W(t)$ is also a doubly stochastic matrix.} with positive diagonal entries, i.e., $\forall i,j\in\scr{N}$,
	$$w_{ii}(t)>0, \; w_{ij}(t)=w_{ji}(t)\geq 0, \; \text{and}$$
	$$\sum_{k=1}^n w_{kj}(t)=\sum_{k=1}^n w_{ik}(t)=1~.$$
	
	A.2) $W(t)$ is consistent with the network connectivity constraint, i.e., if $(i,j)\notin \scr{E}(t)$, then $w_{ij}(t)=0$.
	\hfill $\Box$
\end{assumption} 

Note that Assumption \ref{assum:Doubly_Sto} is a standard one for the conventional distributed averaging problem (without any sequential observations) \cite{Nedic2009b}. 
Such weights $w_{ij}(t)$  can be designed in a distributed manner using the well-known
Metropolis weights \cite{metro2}. 

Moreover, note that  even though   $z_i(t)$  used in the update is a running average over time, 
an agent need not store all its received samples. Instead,
each agent $i$ can only keep track of $z_i(t)$ and update following  
\eq{z_i(t+1)= \frac{tz_i(t)+x_i(t+1)}{t+1}~.\label{equ:ave_z_recur}}

The algorithm \eqref{update}  requires that the communication between any pair of neighboring agents
be bidirectional. Such a requirement may not always be satisfied in applications. For example,
different agents may have distinct  transmission radii.
In this subsection, we introduce  another algorithm to handle the case when the neighbor graph $\bbb{N}(t)$  is directed,
i.e., the communication between agents is  unidirectional. The algorithm makes use of the idea of
the push-sum protocol \cite{kempe}, which solves the conventional distributed averaging problem
for directed neighbor graphs.

Now, we consider the following algorithm (for directed neighbor graphs).
Each agent $i$ has control over two real-valued variables $y_i(t)$ and $v_i(t)$, 
which are initialized as $y_i(1)=x_i(1)$ and $v_i(1)=1$,  respectively.
At each time $t> 1$, each agent $i$ sends the weighted current values
$\frac{y_i(t)}{1+o_i(t)}$ and $\frac{v_i(t)}{1+o_i(t)}$ to all its current neighbors
and updates its variables according to the rules
\begin{align}
\kai{\mu_i}(t+1) &= \frac{y_i(t)}{1+o_i(t)} +\sum_{j\in\scr{N}_i(t)}\frac{y_j(t)}{1+o_j(t)}~, \nonumber\\
v_i(t+1) &=\frac{v_i(t)}{1+o_i(t)}+\sum_{j\in\scr{N}_i(t)}\frac{v_j(t)}{1+o_j(t)}~, \nonumber \\
y_i(t+1) &=\kai{\mu_i}(t+1)+z_i(t+1)-z_i(t)~,\label{update_direct}
\end{align}
where  $o_i(t)$ denotes the number of observers at agent $i$ at time $t$, \kai{and $z_i(t)$ is the running average as defined in \eqref{equ:ave_z}. The quotient ${\kai{\mu_i}(t)}/{v_i(t)}$ can then be shown to converge to $\bar{x}$.}

\subsection{Convergence Results}
To establish  convergence results, 
we first introduce some concepts on the connectivity of  time-varying graphs.
An undirected graph $\bbb{G}$ is called {\em connected} if there is a path between each pair of distinct vertices in $\bbb{G}$.
A directed graph $\bbb{G}$ is called {\em strongly connected} if there is a directed path between each ordered pair of distinct vertices in $\bbb{G}$. 
By the {\em union} of a finite sequence of undirected (or directed) graphs, $\bbb{G}_1,\bbb{G}_2,\ldots,\bbb{G}_p$,
each with the vertex set $\scr{V}$, is meant the undirected (or directed) graph $\bbb{G}$ with vertex set $\scr{V}$
and edge set equaling the union of the edge sets of all  the graphs in the sequence.
We say that such a finite sequence is {\em jointly connected} (or {\em jointly strongly connected}) if the union of its members
is a connected (or strongly connected) graph. We say that an infinite sequence of undirected (or directed) graphs $\bbb{G}_1,\bbb{G}_2,\ldots$
is {\em repeatedly jointly connected} (or {\em repeatedly jointly strongly connected}) if there is a positive integer $r$ such that for each $k\geq 0$,
the finite sequence
$\bbb{G}_{rk+1},\bbb{G}_{rk+2},\ldots,\bbb{G}_{r(k+1)}$  is jointly connected (or jointly strongly connected).


Now we are ready to present the connectivity assumption on the undirected neighbor graphs $\{\bbb{N}(t)\}$, building upon which we establish the  convergence of system \eqref{state}.

\begin{assumption}\label{assum:Joint_Con}
The sequence of undirected neighbor graphs $\{\bbb{N}(t)\}$ is repeatedly jointly connected.
\hfill $\Box$
\end{assumption}

\begin{theorem}\label{thm:convergence}
Let all $n$ agents adhere to the update rule \rep{update}.
Suppose that Assumptions \ref{assum:Doubly_Sto} and \ref{assum:Joint_Con} hold. Then,
\begin{align}
\lim_{t\rightarrow \infty} y_i(t) = \bar{x}~, \;\;\;\;\; \forall i,\label{xxx}
\end{align}
\kai{almost surely. {\color{black}Moreover, the sequence $\{y_i(t)\}$ reaches consensus for all $i\in\mathcal{N}$ with the order of $O({1}/{t})$ w.h.p.}\hfill $\Box$}
\end{theorem}


A precise  proof of the theorem is relegated to  Section \ref{proof}, but here we provide the basic intuition behind it. First, it  is straightforward to verify that
\eq{\lim_{t\rightarrow\infty}\frac{1}{n}\sum_{i=1}^n z_i(t) = \frac{1}{n}\sum_{i=1}^n \bar x_i\label{avg}~.}
Since $y(1)=x(1)=z(1)$, from \rep{state},
it follows that
$\mathbf{1}^\top y(2) = \mathbf{1}^\top z(2)$, where $\mathbf{1}$ denotes the vector whose entries all equal to $1$
and $\mathbf{1}^\top$ denotes its transpose.
By induction, it follows that $\mathbf{1}^\top y(t) = \mathbf{1}^\top z(t)$
for any $t$, which  implies 
by ignoring the small perturbation terms $z(t+1)-z(t)$ (later we will show that this term converges {polynomially} fast to the zero vector $\mathbf{0}$ w.h.p.), $y(t+1) = W(t)y(t)$ leads all $y_i(t)$ to the same value \cite{Nedic2009b}.
Note that, from \rep{avg} and \rep{bar}, $\sum_{i=1}^n z_i(t)$ will converge to $n\bar{x}$, and therefore we could expect that each $y_i(t)$ will converge to $\bar{x}$
since $\mathbf{1}^\top y(t) = \mathbf{1}^\top z(t)$. 

%



	Likewise, we impose the following assumption on the connectivity of the neighbor graph  $\bbb{N}(t)$ when it is directed. 

	\begin{assumption}\label{assum:Joint_Con_Strong}
		The sequence of directed neighbor graphs $\{\bbb{N}(t)\}$ is repeatedly jointly strongly connected.
		\hfill $\Box$
	\end{assumption}

Now we are ready to present the convergence result for the update rule \eqref{update_direct} for directed neighbor graphs, which shows that it also achieves the same convergence result as update rule \eqref{state} for undirected neighbor graphs.

\begin{theorem}\label{thm:converge_DG}
Let all $n$ agents adhere to the update rule \rep{update_direct}.
Suppose that Assumptions \ref{assum:Doubly_Sto} and \ref{assum:Joint_Con_Strong} hold. Then,
\begin{align}
\lim_{t\rightarrow \infty} \frac{\kai{\mu_i}(t)}{v_i(t)} = \bar{x}~, \;\;\;\;\; \forall i~, \label{yyy}
\end{align}
\kai{a.s. {\color{black}Moreover, the sequence $\{{{\mu_i}(t)}/{v_i(t)}\}$ reaches consensus for all $i\in\mathcal{N}$ with the order of $O({1}/{t})$ w.h.p.}\hfill $\Box$}
\end{theorem}

Proof of this theorem is given later in Section \ref{proof}.

\kai{
Based on Theorems \ref{thm:convergence} and \ref{thm:converge_DG}, we also obtain the following corollary which  quantifies the convergence performance of the update rules  for a finite time $t$. The proof is deferred to Section \ref{proof}.}

\begin{corollary}\label{cor:convergence}
	\kai{For  the update   \rep{update}, there exist constants $\lambda_1\in[0,1)$ and $C_1,C_2>0$, such that  after $t\geq (2C_2)^{-1}\cdot\log\left(\frac{2}{\delta(1-2e^{-2C_2})}\right)
	$ iterations, 
	\begin{align*}
	|y_i(t)-\bar{x}|\leq C_1\lambda_1^t+({C_2+K+\bar{x}_i})/{t},~~\forall i\in\scr{N}~~
	\end{align*}
	holds with probability at least $1-\delta$. The same finite-time performance also holds\footnote{Note that the constants $\lambda_1,C_1,C_2$ for the results of updates \eqref{update} and \eqref{update_direct} may be different.} for the update \eqref{update_direct}.}
\end{corollary}

\section{Quantization Effects}\label{sec:quan}

In this section, we investigate the effects of two common sources of quantization on algorithm~\rep{update}. 
Specifically, 
the quantization includes  quantized communication and  limited precision of the division operation, which are common 
in realistic distributed computation systems. The joint effects of the two quantization are considered over both static and dynamic neighbor graphs.

\subsection{Static Graphs}
The algorithm \eqref{update} requires the agents to send and receive real-valued variables $y(t)$ at each iteration. However, in a realistic network, with limited capacity of communication links, the messages transmitted among agents can have only limited length. On the other hand, both  updates   \eqref{equ:ave_z} and \eqref{equ:ave_z_recur} require one step of division operation. As such, the result $z(t)$ may only have limited precision due to the finite digits of representing divided numbers in realistic digital computers. This effect can be viewed as one type of quantization on the value of $z(t)$. With these two quantization effects, the precise belief averaging cannot be achieved in general (except in some special cases). We first  analyze the performance of system \eqref{state} subject to these effects over a static neighbor graph. 
To this end, 
we impose the following standard assumption on the static graph and the corresponding weight matrix.

\begin{assumption}\label{assum:quan_weight}
The communication graph $\bbb{N}(t)$ is static and connected, i.e., $\bbb{N}(t)=(\scr{N},\scr{E})$ for all $t$ where $(\scr{N},\scr{E})$ is a connected undirected graph. Accordingly, there exists a matrix $W$   
which satisfies Assumption \ref{assum:Doubly_Sto} in that $W(t)=W$ for all $t$, and has the following properties: 

	A.1) $W$ has dominant diagonal entries, i.e., $w_{ii}>1/2$ for all $i\in\scr{N}$.
	
	A.2) For any $(i,j)\in\scr{E}$, we have $w_{ij}\in\bbb{Q}^+$, where $\bbb{Q}^+$ is the set of rational numbers in the interval $(0,1)$.
\hfill $\Box$	
\end{assumption}

Note that conditions A.1) and A.2) are specifically needed for the convergence of quantized systems as in \cite{chamie}. 
\kai{
In a practical implementation, 
it is not restrictive to have rational numbers as weights and require dominant diagonal weights. 
As reported in \cite[Appendix A]{chamie}, the quantized update may fail to converge even in the deterministic setting of   distributed averaging. Thus, conditions A.1) and A.2) are essential here since our analysis will rely on the results developed in \cite{chamie}.  
}

Let $\tilde{z}_i(t)$ denote the value of $z(t)$ after the division in \eqref{equ:ave_z}, or equivalently \eqref{equ:ave_z_recur}. We impose an assumption on $\tilde{z}_i(t)$ as follows. 
\begin{assumption}\label{assum:limited_prec}
	There exists a precision $\Delta>0$ such that $\tilde{z}_i(t)$ is  multiples of $\Delta$ for  any time $t$ and $i\in\scr{N}$. \hfill $\Box$
\end{assumption}

It follows from 
Assumption \ref{assum:limited_prec}   that the decimal part of $\tilde{z}_i(t)$ can only have a finite number of values.
For notational convenience, we introduce the following definitions.
Define the sets $\scr{B}$ and $\scr{C}$ as,
\begin{align}\label{equ:Belief_set}
&\scr{B}:=\{b:b=k\Delta+\Delta/2, k=0,1,\cdots\},\\
&\scr{C}:=\{b:b=k\Delta, k=0,1,\cdots\}. 
\end{align}  
Let   $\scr{R}:\mathbb{R}\to\mathbb{R}$ be the operation that 
rounds the value to the nearest multiples of $\Delta$, i.e., 
\eq{\scr{R}(x)=\argmin_{b\in\scr{C}}|b-x|.} 
Hence, we have  $\tilde{z}_i(t)=\scr{R}(z_i(t))$. 
In particular, if  $x\in\scr{B}$ for some $k>0$, we define the value of $\scr{R}(x)$ as $\scr{R}(x):=(k+1)\Delta$.
In practice, the value of the quantized $\tilde{z}_i(t)$ can be evaluated by simply keeping track of the summation of all the previous data at time $t$, i.e., define \kai{$s_i(t):=\sum_{\tau\leq t}x_i(\tau)$}, and then calculate $\tilde{z}_i(t)=\scr{R}(s_i(t)/t)$.
In this regard, the imprecision caused by the division operation will not accumulate over time.

To avoid the information loss caused by the  quantized communication, we adopt the following update:
\eq{y(t+1) = W\scr{Q} (y(t))+y(t)-\scr{Q} (y(t))+\Delta \tilde{z}(t+1)~,\label{quan_state}}
where $\scr{Q}:\mathbb{R}^n\to \mathbb{R}^n$ denotes the operation of element-wise  quantization on a vector, and $\Delta \tilde{z}(t+1):=\tilde{z}(t+1)-\tilde{z}(t)$.
The deterministic quantizer $\scr{Q}$ can be either  truncation quantizer $\scr{Q}_t$,  the ceiling  quantizer $\scr{Q}_c$, or the rounding  quantizer $\scr{Q}_r$, which round the values to, respectively,  the nearest lower integer,  the nearest upper integer, or the nearest integer. For more discussion on the types of deterministic quantizers, see \cite[Section IV]{chamie}.
{Without any loss of generality, we analyze the system with a truncation quantizer $\scr{Q}_t$.}

Since $W$ is column stochastic (i.e., $\mathbf{1}^\top W=\mathbf{1}^\top$), from equation \eqref{quan_state}, we have
\begin{align}\label{equ:quan_state_comp}
\mathbf{1}^\top y(t+1)=\mathbf{1}^\top y(t)+\mathbf{1}^\top \big(\tilde{z}(t+1)-\tilde{z}(t)\big).
\end{align}
Hence, the property that $\mathbf{1}^\top y(t) = \mathbf{1}^\top \tilde{z}(t)$
for any $t$ also holds under quantized communication provided $\mathbf{1}^\top y(1) = \mathbf{1}^\top \tilde{z}(1)$. 
In addition, we define $m(t)$ and $M(t)$ as follows:
$$
m(t):=\min_{i\in\scr{N}}\lfloor y_i(t)\rfloor,~~~M(t):=\max_{i\in\scr{N}}\lfloor y_i(t)\rfloor, ~~\forall t\geq 0~,
$$
where $\lfloor\cdot\rfloor$ denotes the floor function.
Let $\alpha_i=1-w_{ii}+\gamma$, and $\gamma>0$ be a sufficiently small positive scalar\footnote{The exact characterization of how $\gamma$ is selected is given in \cite{chamie}.} that guarantees $\alpha_i<0.5$.
As shown in \cite{chamie}, the  value of $\gamma$ is not necessarily known and is only used here for the convenience of analysis. 
Define $\alpha:=\max\limits_{i\in\scr{N}}\alpha_i$, \kai{which is also upper bounded by $0.5$, }
and a value $\bar x_{\scr R}$ 
as
\begin{align}\label{equ:barxo}
	\bar x_{\scr R}:=\frac{1}{n}\sum_{i\in\scr{N}}\scr{R}(\bar{x}_i)~.
\end{align}  
If $\bar{x}_i\notin \scr{B}, \forall i\in\scr{N}$, then the value of $\bar x_{\scr R}$ corresponds to the average of all beliefs with limited precisions.
Note that the difference between $\bar x_{\scr R}$ and the actual average belief $\bar{x}$ is no greater than $\Delta/2$ by definition.
We will show that the quantized system \eqref{quan_state} will converge to the neighborhood of $\bar x_{\scr R}$ \kai{a.s.,} provided  that  $\bar{x}_i\notin \scr{B}, \forall i\in\scr{N}$. 
Formally, we have the following proposition on the  convergence  of system \eqref{quan_state}.




\begin{proposition}\label{prop:quan_conv_static}
Let all $n$ agents adhere to the update rule \rep{quan_state}.
Under Assumptions  \ref{assum:quan_weight} and \ref{assum:limited_prec}, if  $\bar{x}_i\notin \scr{B}, \forall i\in\scr{N}$, then \kai{almost surely} either 
	
		1) the system reaches quantized consensus to the value $\bar x_{\scr R}$ defined in \eqref{equ:barxo}, i.e.,
		$$
	    \left\{
            \begin{array}{ll}
              \lfloor y_i(t)\rfloor=\lfloor y_j(t)\rfloor, \;\forall i,j \in \scr{N}~,\\
              |y_i(t)-\bar x_{\scr R}|<1, \;\forall i\in\scr{N}~,
            \end{array}
        \right.
		$$
		which implies that $|y_i(t)-\bar{x}|<1+{\Delta}/{2}, \forall i\in\scr{N}$, 
		or
				
		2) all $n$ agents' values live in a small neighborhood around   $\bar x_{\scr R}$ 
        in that
	    $$
	    \left\{
            \begin{array}{ll}
              | y_i(t)- y_j(t)|\leq \alpha_i+\alpha_j, \;\forall i,j \in \scr{N}~,\\
              |y_i(t)-\bar x_{\scr R}|\leq 2\alpha, \;\forall i\in\scr{N}~,
            \end{array}
        \right.
        $$
		which implies that $|y_i(t)-\bar{x}|<2\alpha+{\Delta}/{2}, \forall i\in\scr{N}$. \hfill $\Box$
\end{proposition}



\kai{
Proposition \ref{prop:quan_conv_static} states that under the condition that $\bar{x}_i\notin \scr{B}, \forall i\in\scr{N}$, 
system \eqref{quan_state} will either  reach a quantized consensus with error smaller than $1+{\Delta}/{2}$ to the actual average belief $\bar{x}$, or enter a bounded neighborhood of $\bar{x}$ with  size smaller than $2\alpha+{\Delta}/{2}$. This  result  can be viewed as extension of the one for standard distributed averaging with quantized communications \cite{chamie}. Notably, the quantization effect caused by the division operation  enlarges the error away from exact consensus by an amount of ${\Delta}/{2}$, which is usually small in practice. 
}
The proof of the proposition is provided in Section~\ref{proof}.

\begin{remark}
It is worth noting that the limiting behavior of the quantized system \rep{quan_state} differs from the results in \cite{chamie} in three ways: i) quantized communication does not necessarily cause exact cyclic behavior of $y(t)$ due to  the randomness in the sequential  data sample $x_i(t)$; ii) limited precision of $\tilde{z}(t)$ induces inevitable mismatch to the convergent point  from the actual average belief $\bar{x}$ by a small amount; instead, the system can only converge to  some value $\bar x_{\scr R}$ close to $\bar{x}$ up to a small deviation; iii) the convergence result holds \kai{almost surely}, instead of in deterministic finite number of iterations. 
\hfill $\Box$
\end{remark}

\subsection{Dynamic Graphs}
In this subsection, we extend the previous convergence result over static graphs to dynamic graphs.
It follows from \cite{el2016distributed} that even without sequential data samples, i.e., $\Delta \tilde{z}(t+1)\equiv 0$, there exist counterexamples which show that  quantized communication could prevent   consensus update from converging for  general dynamic graphs.
Therefore, we consider a  special class of dynamic graphs, namely, the probabilistic dynamic graphs model,  where  each link has a positive probability to appear in the graph at any time $t$. 
The probabilistic model is formally detailed in the following assumption. We note that this is still a fairly large class of graphs.

\begin{assumption}\label{assum:dyna_graph_value}
	The neighbor graph $\bbb{N}(t)$ is dynamic. Specifically, 	there exists an underlying graph $(\scr{N},\scr{E})$ and a corresponding  matrix $W$ satisfying Assumption~\ref{assum:Doubly_Sto} and  Assumption~\ref{assum:quan_weight}. 
At each time $t$, $W(t)$ is constructed from $W$ as follows:
	$$
	w_{ij}(t)=\left\{
	\begin{array}{ll}
	w_{ij}, &\text{if~~} (i,j)\in\scr{E}(t)~,\\
	0, &\text{if~~} (i,j)\notin\scr{E}(t)~,
	\end{array}
	\right.
	$$
	and $w_{ii}(t)=1-\sum_{j\in\scr{N}_i}w_{ij}(t)$. Moreover, let $\{\scr{F}_t\}_{t\geq 1}$ be the $\sigma$-field  generated by the random graphs $\{\bbb{N}(\tau)\}_{\tau\leq t}$, i.e., $\scr{F}_t=\sigma(\bbb{N}(1),\cdots,\bbb{N}(t))$. Then, $Prob[(i,j)\in{\color{black}\scr{E}(t)}|\scr{F}_{t-1}]\geq p$ for all $t$ and all $(i,j)\in\scr{E}$, where $p>0$ is a positive constant and $(\scr{N},\scr{E})$ is a connected undirected graph.\hfill $\Box$
\end{assumption}

\kai{We note that Assumption \ref{assum:dyna_graph_value},  a probabilistic model for dynamic graphs, is different from the deterministic  models  in Assumptions \ref{assum:Joint_Con} and \ref{assum:Joint_Con_Strong}.} 
This probabilistic  model has been  adopted in many prior work on distributed averaging, including asynchronous gossiping graphs \cite{lavaei2012quantized}, wireless sensor networks subject to probabilistic link failures \cite{patterson2010convergence},
and conventional distributed averaging with quantized communication \cite{el2016distributed}.

The update rule over the dynamic graph thus becomes
\eq{y(t+1) = {W}(t)\scr{Q} (y(t))+y(t)-\scr{Q} (y(t))+\Delta \tilde{z}(t+1)~.\label{quan_state_dyna}}
The following proposition describes  the limiting behavior of system \eqref{quan_state_dyna}, the proof of which is provided in Section \ref{proof}. 

\begin{proposition}\label{prop:quan_conv_dynamic}
Let all $n$ agents adhere to the update rule \rep{quan_state_dyna}.
Under Assumption  \ref{assum:dyna_graph_value}, if  $\bar{x}_i\notin \scr{B}, \forall i\in\scr{N}$, then the result in Proposition \ref{prop:quan_conv_static} still holds.\hfill $\Box$
\end{proposition}

\section{Analysis}\label{proof}

In this section, we provide proofs of the results presented in Sections \ref{sec:static} and \ref{sec:quan}.

\subsection{Preliminaries}
The proofs for the results in Section \ref{sec:static} will appeal to the stability properties of discrete-time
linear consensus processes.
We begin with the idea of a certain semi-norm which was introduced in \cite{pieee}. 
 Let $||\cdot ||$ be the induced infinity norm on $\R^{m\times n}$.
For $M\in\R^{m\times n}$, define
$$|M|_{\infty} = \min_{c\in\R^{1\times n}}||M-\mathbf{1}c||~.$$
It has been shown in \cite{pieee} that $|\cdot|_{\infty}$ is a {\em semi-norm}, namely that it is positively homogeneous
and satisfies the triangle inequality.
Moreover, this particular semi-norm is {\em sub-multiplicative} 
(see Lemma 1 in \cite{cdc14}).
\kai{In particular, from Lemmas 2 and 3 in \cite{cdc14}, for any $x\in\R^n$ and nonnegative matrix $A\in\R^{n\times n}$, we have 
\begin{align}\label{equ:inf1}
	|x|_{\infty} &= \frac{1}{2}\displaystyle\left(\max_{i}x_i-\min_{j}x_j\right)~\\
	 |A|_{\infty} &= \frac{1}{2}\max_{i,j}\sum_{k=1}^{n}|a_{ik}-a_{jk}|~.\notag
\end{align}
}
Moreover, from \cite{seneta},  $|A|_{\infty}\le 1$ if $A$ is a stochastic matrix.

{\color{black}
We first introduce the notion of internal stability, which has been proposed and studied in \cite{cdc14}.

Consider a discrete-time linear consensus process modeled by a linear recursion equation of the form
\eq{x(k+1)=S(k)x(k)~, \;\;\;\;\; x(k_0)=x_0~,\label{con}}
where $x(k)$ is a vector in $\R^n$ and $S(k)$ is an $n\times n$ stochastic matrix.
It is easy to verify  that the equilibria of \rep{con} include points of the form $a\mathbf{1}$.
We say that the system described by \rep{con} is {\em uniformly exponentially consensus stable} if
there exist a finite positive constant $\gamma$  and a constant $0\leq \lambda<1$ such that
for any $k_0$ and $x_0$, the corresponding solution satisfies
$$|x(k)|_{\infty}\leq \gamma \lambda^{k-k_0}|x_0|_{\infty}~, \;\;\;\;\; k\geq k_0~.$$
Uniform exponential consensus stability implies that  solutions of \rep{con} approach a consensus vector (i.e., all the entries of $x(t)$ have the same value) exponentially fast.



Exponential consensus stability can be characterized by graph connectivity.
Toward this end, we need the following concept.
The {\em graph of a nonnegative symmetric matrix} $M\in\R^{n\times n}$, denoted by $\gamma(M)$, is an
undirected graph on $n$ vertices with an
edge between vertex $i$ and vertex $j$ if and only if $m_{ji}\neq 0$ (and thus $m_{ij}\neq 0$).


\begin{lemma}
Let $\scr{F}$ denote a compact subset of the set of all $n\times n$ symmetric stochastic matrices
with positive diagonal entries. Suppose that $F(1),F(2),\ldots$ is an infinite sequence of matrices in $\scr{F}$.
Then, the discrete-time linear recursion equation
$x(k+1)=F(k)x(k)$, $k\geq 1$,
is uniformly exponentially consensus stable if and only if the sequence of graphs
$\gamma(F(1)),\gamma(F(2)),\gamma(F(3))\ldots$
is repeatedly jointly connected.
\hfill $\Box$
\label{nec}\end{lemma}


This lemma is a direct consequence of Theorem 4 in \cite{cdc14}.


Now we turn to
input-output stability of discrete-time linear consensus processes.
Toward this end,
we rewrite the equation \rep{con} in an input-output form as follows:
\begin{align}
x(k+1) &= S(k)x(k) + B(k)u(k)~, \label{in}\\
y(k) &= C(k)x(k)~.
\label{inout}\end{align}
We are interested in the case when $B(k)$ and $C(k)$ are stochastic matrices for all $k$.
We say that the system defined by \rep{in}-\rep{inout} is {\em uniformly bounded-input,
bounded-output consensus stable} if there exists a finite constant $\eta$ such that for any $k_0$ and any
input signal $u(k)$ the corresponding zero-state response satisfies
$$\sup_{k\geq k_0} |y(k)|_{\infty} \leq \eta \sup_{k\geq k_0} |u(k)|_{\infty}~.$$
It is worth noting that $y(t)$ may not be bounded even though the system is
uniformly bounded-input, bounded-output consensus stable.

The following result establishes the connection between uniform bounded-input, bounded-output
stability, and uniform exponential stability.


\begin{proposition}
{\rm (Theorem 2 in \cite{cdc16})}
Suppose that \rep{con} is uniformly exponentially consensus stable.
Then, the system  \rep{in}-\rep{inout}  is uniformly bounded-input, bounded-output consensus stable.\hfill $\Box$
\label{out}\end{proposition}
}



\subsection{Proof of Theorem \ref{thm:convergence}}



{The system \rep{state} can be viewed as a linear consensus system with input
$u(t)=z(t+1)-z(t)$, $S(t)=W(t)$ a stochastic matrix, and $B(t)=C(t)=I$, where $I$
is the identity matrix, which is also a stochastic matrix. 
Let $\Phi(t,\tau)$ be the
discrete-time state transition matrix of $S(t)$, i.e.,
\begin{align}\label{equ:Def_Phi}
\Phi(t,\tau) = \left\{ \begin{array}{ll}
  S(t-1)S(t-2)\cdots S(\tau) &\mbox{ if  $t>\tau$}~, \\
  I &\mbox{ if  $t=\tau$}~.
       \end{array} \right.
\end{align}
It is easy to verify that $\Phi(t,\tau)$ is a stochastic matrix for any $t\geq \tau$. 
Then, the output  is given by 
$y(t)= \Phi(t,1)y(1) + \hat y(t)$, 
where $\hat y(t)$ is the zero-state response and the first component on the right-hand side
is the zero-input response.
It then follows that
\begin{eqnarray}\label{equ:yt_response_bnd}
|y(t)|_{\infty} &\le& |\Phi(t,1)y(1)|_{\infty} + |\hat y(t)|_{\infty} \notag\\
&\le& |\Phi(t,1)y(1)|_{\infty} + \sup_{\tau\ge t}|\hat y(\tau)|_{\infty}~.
\end{eqnarray}
Since the sequence of neighbor graphs is repeatedly jointly connected,
by Lemma \ref{nec}, the system $y(t+1)=W(t)y(t)$ is uniformly exponentially consensus stable,
and thus the system \rep{state} is uniformly input-bounded, output-bounded consensus stable by Proposition~\ref{out}.
Since $|\Phi(t,1)y(1)|_{\infty}$ converges to $0$ exponentially fast,
{\color{black} to prove the theorem, it suffices to show that $\sup_{\tau\geq t} |\hat y(\tau)|_{\infty}$ converges to $0$ almost surely  with the order of $O({1}/{t})$ with high probability, and the consensual value reached by the sequence $\{y_i(t)\}$ is indeed $\bar{x}$. }
Since $\sup_{\tau\geq t} |\hat y(\tau)|_{\infty}\le \eta\sup_{\tau\geq t} |u(\tau)|_{\infty}$ for some constant $\eta$, and noting that from \eqref{equ:inf1}, 
$\sup_{\tau\geq t} |u(\tau)|_{\infty}\le 2\sup_{\tau\geq t} \max_i |u_i(\tau)|$,
it will be enough  to study the convergence  of $\sup_{\tau\geq t} \max_i|u_i(\tau)|$. 

Note that $u_i(t) = z_i(t+1)-z_i(t)$ for any $i\in\scr{N}$, and
\begin{eqnarray*}
&&\left|z_i(t+1)-z_i(t)\right| = \left|\frac{\sum_{k=1}^{t+1}x_i(k)}{t+1}-\frac{\sum_{k=1}^{t}x_i(k)}{t}\right|\\
&=& \frac{\left|t\cdot x_i(t+1)-\sum_{k=1}^t x_i(k)\right|}{t(t+1)}\\
&\leq& \frac{ \left|x_i(t+1)\right|}{t+1}+\frac{\left|\sum_{k=1}^t x_i(k)\right|}{t(t+1)}~.
\end{eqnarray*}
\kai{
Recall that $|X_i(\omega)|\leq K$ for some constant $K$. Also, by the Strong Law of Large Numbers, we have that ${|\sum_{k=1}^t x_i(k)|}/{t}$ converges to $\bar{x}_i$ a.s. Thus, ${|\sum_{k=1}^t x_i(k)|}/{[t(t+1)]}$ converges to $0$ a.s., which implies {\color{black} that almost surely, the sequence $\{y_i(t)\}$ reaches to a consensual value $\bar{y}$ for all $i\in\mathcal{N}$. Moreover, since $\mathbf{1}^\top y(t) = \mathbf{1}^\top z(t)$ and thus $\sum_{i\in\mathcal{N}}y_i(t)/n=\sum_{i\in\mathcal{N}}z_i(t)/n$ holds for any $t\geq 0$,  and also $z_i(t)$ converges to $\bar{x}_i$ a.s. (by the Strong Law of Large Numbers), we have $\sum_{i\in\mathcal{N}}y_i(t)/n$ converges to $\sum_{i\in\mathcal{N}}\bar{x}_i/n=\bar{x}$ a.s. Therefore, we obtain that the consensual value $\bar{y}$ equals the value of $\bar{x}$, which concludes the first argument in Theorem \ref{thm:convergence}. } 
}

In addition, 
using the Chernoff bound, for any $\delta >0$, with a probability of at least $1-2e^{-2\delta t}$, we have
$$
\left|\frac{\sum_{k=1}^t x_i(k)}{t}-\bar{x}_i\right| \leq \delta~.
$$
Then,
\begin{eqnarray*}
&& P\left(\sup_{\tau \geq t}\left|\frac{\sum_{k=1}^{\tau} x_i(k)}{\tau}-\bar{x}_i\right| \leq \delta\right)\\
&=& P\left(\forall \tau \geq t,\; \left|\frac{\sum_{k=1}^{\tau} x_i(k)}{\tau}-\bar{x}_i\right| \leq \delta\right)\\
&\geq& 1 - \sum_{\tau \geq t} P\left(\left|\frac{\sum_{k=1}^{\tau} x_i(k)}{\tau}-\bar{x}_i\right| > \delta\right)\\
&\geq& 1- \sum_{\tau \geq t} 2e^{-2\delta \tau} =1-\frac{2e^{-2\delta t}}{1-e^{-2\delta}}~,
\end{eqnarray*}
which implies that w.h.p.,
\eq{
\sup_{\tau \geq t} \left|\frac{\sum_{k=1}^{\tau} x_i(k)}{\tau}\right| \leq \bar{x}_i +\delta~,
\label{equ:sup_zt}}
and furthermore, 
\eq{
\sup_{\tau \geq t} \left|z_i(\tau+1)-z_i(\tau)\right| \leq \frac{K+\bar{x}_i+\delta}{t+1}~,
\label{equ:sup_diffzt}}
which is decreasing uniformly with  the order of $O({1}/{t})$.
{\textcolor{black}{From \eqref{equ:yt_response_bnd}, this means that the sequence $\{y(t)\}$ reaches consensus with the rate of $O({1}/{t})$, which completes the proof.}}
\hfill$\qed$

\subsection{Proof of Theorem \ref{thm:converge_DG}}
The proof of  Theorem \ref{thm:converge_DG} also relies on the concept of uniformly exponentially consensus stability and Proposition~\ref{out}. 
Define the state in update \eqref{xxx} as
\begin{align}\label{equ:Def_h_i}
	h_i(t):=\frac{\kai{\mu_i}(t)}{v_i(t)},~~\forall i\in\scr{N}.
\end{align}
For notational convenience, we also define, for any time $t$,
$$
\Delta z_i(t):=z_i(t)-z_i(t-1),~~l_i(t):=\frac{1}{1+o_i(t)},~~\forall i\in\scr{N}.
$$
We first rewrite the update \eqref{update_direct} as an input-output system of the form  \eqref{in}-\eqref{inout}. In particular, from \eqref{update_direct}, we have
\begin{align}
	&h_i(t+1)=\frac{\sum_{j\in\scr{N}^+_i(t)}l_j(t)(w_j(t)+\Delta z_j(t))}{\sum_{j\in\scr{N}^+_i(t)}l_j(t)v_j(t)}\nonumber\\
	&=h_i(t)\cdot \frac{l_i(t)\left(1+{\Delta z_i(t)}/{\kai{\mu_i}(t)}\right)}{l_i(t)+\sum_{j\in\scr{N}_i(t)}l_j(t)\cdot {v_j(t)}/{v_i(t)}}\nonumber\\
	&+\sum_{j\in\scr{N}_i(t)}h_j(t)\cdot \frac{l_j(t)\left(1+{\Delta z_j(t)}/{w_j(t)}\right)}{l_j(t)+\sum_{k\in\scr{N}^+_i(t),k\neq j}l_k(t)\cdot{v_k(t)}/{v_j(t)}}~,\nonumber
\end{align}
where we recall that $\scr{N}^+_i(t)$ is the set of neighbors including agent $i$ at time $t$.
Let the entries of $S(t)$ and $B(t)$ in \eqref{in} be
\begin{align}\label{equ:Def_s_ij}
	s_{ij}(t)=\frac{l_j(t)v_j(t)}{\sum_{j\in\scr{N}^+_i(t)}l_j(t){v_j(t)}}, ~~ 
	b_{ij}(t)=s_{ij}(t),
\end{align}
respectively. Then, the update \eqref{update_direct} can be written as 
\begin{align}\label{equ:h_state}
	h(t+1)=S(t)h(t)+B(t)\left(\Delta z(t)\oslash v(t)\right),
\end{align}
where $\oslash$ denotes the Hadamard (element-wise) division operation.

Note that both $S(t)$ and $B(t)$ are stochastic matrices. 
The following lemma shows the uniformly exponentially consensus stability of the zero-input response (i.e., the input $\Delta z(t)=0$).

\begin{lemma}\label{lemma:uniexstable_DG}
	Under Assumptions \ref{assum:Doubly_Sto} and \ref{assum:Joint_Con_Strong}, the system $h(t+1)=S(t)h(t)$ is uniformly exponentially consensus stable, with $h(t)$ and $S(t)$ defined in \eqref{equ:Def_h_i} and \eqref{equ:Def_s_ij},  respectively.\hfill $\Box$
\end{lemma}

{\em Proof:}
From Lemma 1 (a) in \cite{nedic2015distributed}, we know that under Assumptions \ref{assum:Doubly_Sto} and \ref{assum:Joint_Con_Strong},  there exist  constants $C'>0$ and $\lambda\in(0,1)$ such that for any $i\in\scr{N}$
$$
\bigg|h_i(t+1)-\frac{\mathbf{1}^\top y(t)}{n}\bigg|\leq C'\cdot \lambda ^t.
$$
Hence, there exists a constant $C=C'/(\lambda\cdot|h(0)|_{\infty})>0$ such that
\begin{align}
|h(t)|_{\infty}=&\frac{1}{2}\left(\max_{i}h_i(t)-\min_{j}h_j(t)\right)\nonumber\\
\leq &\frac{1}{2}\bigg|\max_{i}h_i(t)-\frac{\mathbf{1}^\top y(t-1)}{n}\bigg|\nonumber\\
&+\frac{1}{2}\bigg|\max_{j}h_j(t)-\frac{\mathbf{1}^\top y(t-1)}{n}\bigg|\nonumber\\
\leq &\frac{C'}{\lambda}\cdot \lambda^t= C \cdot |h(0)|_{\infty}\cdot \lambda^t.
\end{align}
By definition, the update is uniformly exponentially consensus stable, which completes the proof.
\hfill \qed

We are now in a position to prove Theorem \ref{thm:converge_DG}.
 
{\em Proof of Theorem \ref{thm:converge_DG}:}
The system \rep{equ:h_state} can be viewed as a linear consensus system with input
$u(t)=\Delta z(t)\oslash v(t)$ and $C(t)=I$.
Recall the definition of $\Phi(t,\tau)$ in \eqref{equ:Def_Phi}, we can write the  output as 
$$
h(t)= \Phi(t,1)h(1) + \hat h(t)~,
$$
where $\Phi(t,1)$ is as defined in \eqref{equ:Def_Phi}, $\hat h(t)$ and $\Phi(t,1)h(1)$  are the zero-state and zero-input responses, respectively.
Since
$$
|h(t)|_{\infty} \le |\Phi(t,1)h(1)|_{\infty} + \sup_{\tau\ge t}|\hat h(\tau)|_{\infty}~,
$$
and $|\Phi(t,1)h(1)|_{\infty}$ converges to zero exponentially fast according to Lemma \ref{lemma:uniexstable_DG}, it suffices to study the convergence rate of $\sup_{\tau\ge t}|\hat h(\tau)|_{\infty}$.

In addition, by Proposition \ref{out} and Lemma \ref{lemma:uniexstable_DG}, the update \eqref{equ:h_state} is uniformly input-bounded, output-bounded consensus stable. Hence, it follows that $\sup_{\tau\geq t} |\hat y(\tau)|_{\infty}\le \eta\sup_{\tau\geq t} |u(\tau)|_{\infty}$ for some constant $\eta>0$. It is thus sufficient to bound the convergence rate of $\sup_{\tau\geq t} \max_i|\Delta z_i(\tau)/v_i(\tau)|$.
By Lemma~3 in \cite{acc12}, there exists a constant $\epsilon>0$, such that for any $i\in\scr{N}$, the state $v_i(t)$ that follows the update \eqref{xxx} is lower bounded by $\epsilon$.
Thus we obtain
\begin{align}\label{equ:direc_to_zt}
\bigg|\frac{\Delta z_i(t)}{v_i(t)}\bigg|\leq \frac{1}{\epsilon}\cdot\left|z_i(t)-z_i(t-1)\right|.	
\end{align}
From the proof of Theorem \ref{thm:convergence}, \kai{this implies that i) $|\Delta z_i(\tau)/v_i(\tau)|$  converges to zero a.s.; ii) $|\Delta z_i(\tau)/v_i(\tau)|$ also converges  with a  rate of $O({1}/{t})$ w.h.p., which completes the proof.} 
\hfill \qed 

\kai{
\subsection{Proof of Corollary  \ref{cor:convergence}}
From \eqref{equ:yt_response_bnd}, we  further obtain that
\begin{align*}
|y(t)|_{\infty}\le |\Phi(t,1)y(1)|_{\infty} + \sup_{\tau\ge t}| z_i(\tau+1)-z_i(\tau)|~,
\end{align*}
for some constant $\eta>0$.
Since $\Phi(t,1)$ is a stochastic matrix, there exists a $\lambda_1\in[0,1)$ and $C_1>0$ such that $|\Phi(t,1)y(1)|_{\infty}\leq C_1\lambda_1^t$. Also, from \eqref{equ:sup_diffzt}, we know that there exists $C_2>0$ such that  $\sup_{\tau \geq t} \left|z_i(\tau+1)-z_i(\tau)\right|\leq C_2/t$ with  probability at least $1-[{2e^{-2(C_2-K-\bar{x}_i) t}}]/[{1-e^{-2(C_2-K-\bar{x}_i)}}]$. Letting  $\delta=[{2e^{-2(C_2-K-\bar{x}_i) t}}]/[{1-e^{-2(C_2-K-\bar{x}_i)}}]$, we obtain the desired expression for $t=t(\delta)$, which completes the proof of the first argument. The second argument also holds due to \eqref{equ:direc_to_zt}, which relates $|{\Delta z_i(t)}/{v_i(t)}|$ to $\left|z_i(t)-z_i(t-1)\right|$. This concludes  the proof.
\hfill \qed 
}

\subsection{Proof of Proposition \ref{prop:quan_conv_static}}\label{subsec:proof_prop_1}
The proofs for the results in Section \ref{sec:quan} will depend on the results in \cite{chamie} and \cite{el2016distributed}.
To prove Proposition \ref{prop:quan_conv_static}, we first state  the following lemma, which is in the spirit of   Proposition 1 in \cite{el2016distributed}.

\begin{lemma}\label{lemma:finite_values}
Consider the quantized system \rep{quan_state}. Under Assumptions \ref{assum:quan_weight} and  \ref{assum:limited_prec}, if  $\bar{x}_i\notin \scr{B}, \forall i\in\scr{N}$, then \kai{almost surely}, the values of $y_i(t)$, $t\in\{1,2,\ldots\}$, belong to a finite set for all $i\in\scr{N}$. \hfill $\Box$
\end{lemma}

{\em Proof:}
As shown in \cite{chamie},  the three types of quantizers from communications, i.e., truncation quantizer $\scr{Q}_t$, ceiling quantizer $\scr{Q}_c$, and rounding quantizer $\scr{Q}_r$, are all related and can be transformed to each other. Thus, it is sufficient to analyze the quantized update \eqref{quan_state} using any one of the types\footnote{Note that the update \eqref{quan_state} using rounding and truncation quantizers are identical, while the update using ceiling quantizer $\scr{Q}$ is slightly different since $\scr{Q}_c(x)=-\scr{Q}_t(-x)$, which changes the sign of the last two terms $\tilde{z}(t+1)-\tilde{z}(t)$. However, as we show in the proof, this difference term does not invalidate the result in Lemma \ref{lemma:finite_values}.}. Without any loss of generality, we focus here on the truncation quantizer, i.e.,  $\scr{Q}(\cdot)=\lfloor \cdot \rfloor$.

Under A.2) in Assumption \ref{assum:quan_weight}, there exist  co-prime positive integers $a_{ij}$ and $b_{ij}$ such that $w_{ij}=a_{ij}/b_{ij}$.
Let $B_i$ be the least common multiple of the integers $\{b_{ij}:\forall j\in\scr{N}_i\}$, where $\scr{N}_i$ is the set of neighbors of agent $i$ on the static graph $(\scr{N},\scr{E})$.
Define the decimal part of $y_i(t)$ as $c_i(t):=y_i(t)-\lfloor y_i(t) \rfloor$. Then, $c_i(t)\in[0,1)$ satisfies
\begin{align}
	c_i(t)=&y_i(t)-\lfloor y_i(t) \rfloor\nonumber\\
	=&\sum_{j\in\scr{N}_i}w_{ij}(\lfloor y_j(t-1) \rfloor-\lfloor y_i(t-1) \rfloor)-\lfloor y_i(t)\rfloor\nonumber\\
	&+y_i(t-1)+\Delta \tilde{z}_i(t)\nonumber\\
	=&\sum_{j\in\scr{N}_i}\frac{a_{ij}}{b_{ij}}(\lfloor y_j(t-1) \rfloor-\lfloor y_i(t-1) \rfloor)-\lfloor y_i(t)\rfloor\nonumber\\
	&+c_i(t-1)+\lfloor y_i(t-1) \rfloor+ \Delta \tilde{z}_i(t)\nonumber\\
	=&c_i(t-1)+\frac{Z_i(t)}{B_i}+\Delta \tilde{z}_i(t),
\end{align}
where $Z_i(t)\in\bbb{Z}$ is an integer. Note that $\Delta \tilde{z}(t)$ can only take values of multiples of $\Delta$. Hence, $c_i(t)\in[0,1)$ can only take a finite number of values.

We first  show that as $t\to\infty$,  $\tilde{z}_i(t+1)-\tilde{z}_i(t)=0$ a.s. In particular, 
let $\scr{R}_h(x)$ denote the operation of  finding the belief in $\scr{B}$ that is  closest to $x$, i.e., 
\eq{\scr{R}_h(x)=\argmin_{b\in\scr{B}}|b-x|.}
Recall the definition of the belief set $\scr{B}$ in \eqref{equ:Belief_set}. Since $\bar{x}_i\notin \scr{B}$, we have $|\bar{x}_i-\scr{R}_h( \bar{x}_i)|>0, \forall i\in\scr{N}$.
\kai{By the Strong Law of Large Numbers, we have $\lim_{t\to\infty}\left|z_i(t)-\bar{x}_i\right|=0$ a.s. Thus, for any sample realization $\omega$, let $\delta=|\bar{x}_i-\scr{R}_h( \bar{x}_i)|$; then there exists an integer $T_{i}$ such that $\left|z_i(t)(\omega)-\bar{x}_i\right|\leq \delta$ for any $t\geq T_i$. In addition, for any $x\in\mathbb{R}$, if $|x-\bar{x}_i|\leq|\bar{x}_i-\scr{R}_h( \bar{x}_i)|$, then $\scr{R}(x)=\scr{R}(\bar{x}_i)$.
Hence, for any $t\geq T_{i}$, it follows that
$$
\tilde{z}_i(t+1)=\scr{R}({z}_i(t+1))=\scr{R}(\bar{x}_i)=\scr{R}({z}_i(t))=\tilde{z}_i(t),
$$
for this realization $\omega$. }
%
%
Therefore, we obtain
\begin{align*}
	y_i(t+1)=&y_i(t)+\sum_{j\in\scr{N}_i}w_{ij}(\lfloor y_j(t)\rfloor-\lfloor y_i(t)\rfloor)\\
	\leq &c_i(t)+\lfloor y_i(t)\rfloor+\left(\sum_{j\in\scr{N}_i}w_{ij}\right)(M(t)-\lfloor y_i(t)\rfloor)\\
	\leq &c_i(t)+M(t).
\end{align*}
This reduces to the argument of  Proposition 1 in \cite{chamie}, which  implies that $\lfloor y_i(t+1)\rfloor\leq M(t)$, and thus $\{M(t)\}$ is a non-increasing sequence. Similarly, we can show that $\{m(t)\}$ is a non-decreasing sequence. \kai{Therefore, the integer part of $y_i(t)$ (i.e., $\lfloor y_i(t)\rfloor$) takes values in the finite set $\{m(T_i),m(T_i)+1,\cdots,M(T_i)-1,M(T_i)\}$.}
\kai{Since both the integer and the decimal parts  take finite numbers of values, so does  the value of $y(t)$. Note that the argument above holds for any realization $\omega$ in the sample space, which   completes  the proof.}
\hfill$\qed$

Lemma \ref{lemma:finite_values} implies that with sufficiently long time, the limited precision of the  running average of the data    becomes negligible \kai{almost surely} in the analysis of the quantized system \eqref{quan_state}.

We next prove  Proposition \ref{prop:quan_conv_static}.

{\em Proof of Proposition \ref{prop:quan_conv_static}:}
From the proof of Lemma \ref{lemma:finite_values},  if $\bar{x}_i\notin\scr{B}, \forall i$, \kai{for any realization $\omega$, there exists a $T_i$ for each $i\in\scr{N}$, such that  for $t\geq T_i$, the term $\Delta \tilde{z}_i(t)$ will be zero.  Let $T_0=\max_{i\in\scr{N}}T_i$; then from the iteration  $T_0$ on, the system \eqref{quan_state} reduces to  system (11) in \cite{chamie} with initial values $y_i(T_0), \forall i\in\scr{N}$. 
}
Moreover, we have obtained that for any $t\geq T_0$, it  holds that $\tilde{z}_i(t)=\scr{R}(\bar{x}_i)$. Hence, the average of $y(T_0)$ satisfies
$$
\frac{\mathbf{1}^\top y(T_0)}{n}=\frac{\mathbf{1}^\top \tilde{z}(T_0)}{n}=\frac{1}{n}\sum_{i\in\scr{N}}\scr{R}(\bar{x}_i)=\bar x_{\scr R}, 
$$
since ${\mathbf{1}^\top y(T_0)}/{n}={\mathbf{1}^\top \tilde{z}(T_0)}/{n}$ holds for all time $t$.
Then, the statement in Proposition \ref{prop:quan_conv_static} follows directly from Proposition $4$ in \cite{chamie}. Moreover, 
note that the difference between $\bar x_{\scr R}$ and actual average belief  $\bar{x}$ is no greater than $\Delta/2$, since $|\scr{R}(\bar{x}_i)-\bar{x}_i|< \Delta/2, \forall i\in\scr{N}$.  This further bounds the deviation  between $y_i(t)$ and $\bar{x}_i$. \kai{Note that the argument above holds for any realization $\omega$, which completes the proof.}
\hfill$\qed$

\subsection{Proof of Proposition \ref{prop:quan_conv_dynamic}}
The proof of Proposition \ref{prop:quan_conv_dynamic} is similar to that of Proposition~\ref{prop:quan_conv_static}, where the key is to ensure that $y_i(t)$ only takes a finite number of values \kai{a.s.}
We thus first present the following lemma.

\begin{lemma}\label{lemma:finite_values_dynamic}
Consider the quantized system \rep{quan_state_dyna}. 
Under Assumption \ref{assum:dyna_graph_value}, if  $\bar{x}_i\notin \scr{B}, \forall i\in\scr{N}$, then \kai{almost surely,}  the values of $y_i(t)$, $t\in\{1,2,\ldots\}$, belong to a finite set for all $i\in\scr{N}$. \hfill $\Box$
\end{lemma}

{\em Proof:}
\kai{
The proof for that the integer part $\lfloor y_i(t)\rfloor$ can only take a finite number of values  is identical to that  in Section \ref{subsec:proof_prop_1}.
Moreover, by Assumption  \ref{assum:dyna_graph_value}, 
  $w_{ij}(t)$ takes values of  either $w_{ij}=a_{ij}/b_{ij}$ or $0$. 
Thus, for the decimal part, there still exists the least common multiple $B_i'$ for the integers $\{b_{ij}:\forall j\in\scr{N}, j\neq i\}$. Note that by Assumption  \ref{assum:dyna_graph_value},  we need to consider all other agents $j\in\scr{N}$, $j\neq i$, which are possibly connected with agent $i$. Therefore, $c_i(t)=c_i(t-1)+{Z_i(t)}/{B_i'}+\Delta \tilde{z}_i(t)$ still holds for some time-varying integer ${Z_i(t)}$, which implies that $c_i(t)$ can only take a finite number of values. The rest of proof follows directly from  the proof of Lemma  \ref{lemma:finite_values}.
\hfill \qed
}

Similarly, as shown in the proof of Lemma \ref{lemma:finite_values}, the  limited precision of the  running average of the data    becomes negligible a.s.  with sufficiently long time.   Thus, by Lemma \ref{lemma:finite_values_dynamic}, the proof of Proposition \ref{prop:quan_conv_dynamic} follows  from that of Theorem $2$ in \cite{el2016distributed}, and we will not repeat it here for brevity.



\begin{remark}\label{remark:B_set}
Note that for the case when $\bar{x}_i\in\scr{B}$ for some $i\in\scr{N}$, \kai{it is not clear whether the convergence results in \cite{chamie} and \cite{el2016distributed}  can be extended to the setting here with the joint quantization effects.} In this case, the term $\Delta \tilde{z}_i(t)$ becomes a random variable that does not vanish to zero, since the consecutive samples ${z}_i(t)$ and ${z}_i(t+1)$ are drawn around $\bar{x}_i$ and can be  truncated  to either $\bar{x}_i+\Delta/2$ or $\bar{x}_i-\Delta/2$. Therefore, the term $\Delta \tilde{z}_i(t)$ can take values of $\Delta$, $-\Delta$, or $0$, randomly at any time $t$. Based on extensive simulations, we conjecture that the random error $\Delta \tilde{z}_i(t)$ will not accumulate, and that systems \eqref{equ:quan_state_comp} and \eqref{quan_state_dyna} will asymptotically enter a small neighborhood around the desired average belief a.s. We will illustrate this via numerical simulations in Section \ref{sec:simu}, but a formal proof of this result is yet not available.
\hfill $\Box$
\end{remark}

\section{Numerical results}\label{sec:simu}

In this section, we illustrate the convergence performance of the proposed algorithms through numerical examples. We consider a network of $n=10$ agents. Throughout the discussion of our numerical results, the local beliefs of each agent $\bar{x}_i$ are generated uniformly from $[0,100]$, and the sequential samples $\{x_i(t)\}_{t=1}^{\infty}$ are generated from a normal distribution with each local belief $\bar{x}_i$ as the mean and $10$ as the variance.
At any time $t$, we use the average error
$$
e(t) := \frac{\sum_{i=1}^n |y_i(t)-\bar{x}|}{n}
$$
to capture the convergence performance.



\subsection{Undirected Graphs}\label{sec:sim_undirected}

We first study the convergence performance over a static undirected graph.  We test the proposed algorithm on  connected  Random
Geometric Graphs (RGG). The RGGs are generated following \cite{chamie} with connectivity radius $R$ selected as  $R=\sqrt{10*\log(n)/n}=1$. This choice of connectivity radius has been adopted by many in the  literature on RGG \cite{chamie}. The doubly stochastic matrix $W$ is generated following the Metropolis weights \cite{metro2}.  For each fixed $W$, we have repeated the simulation $20$ times (in terms of different realizations of agents' observed samples) and present simulation results on convergence in Fig. \ref{static:1} (with the curve representing average error and shaded region for variance).

\begin{figure}[!ht]
\centering
\subfloat[Convergence of the average error]{\label{static:1}\includegraphics[width=0.45\textwidth]{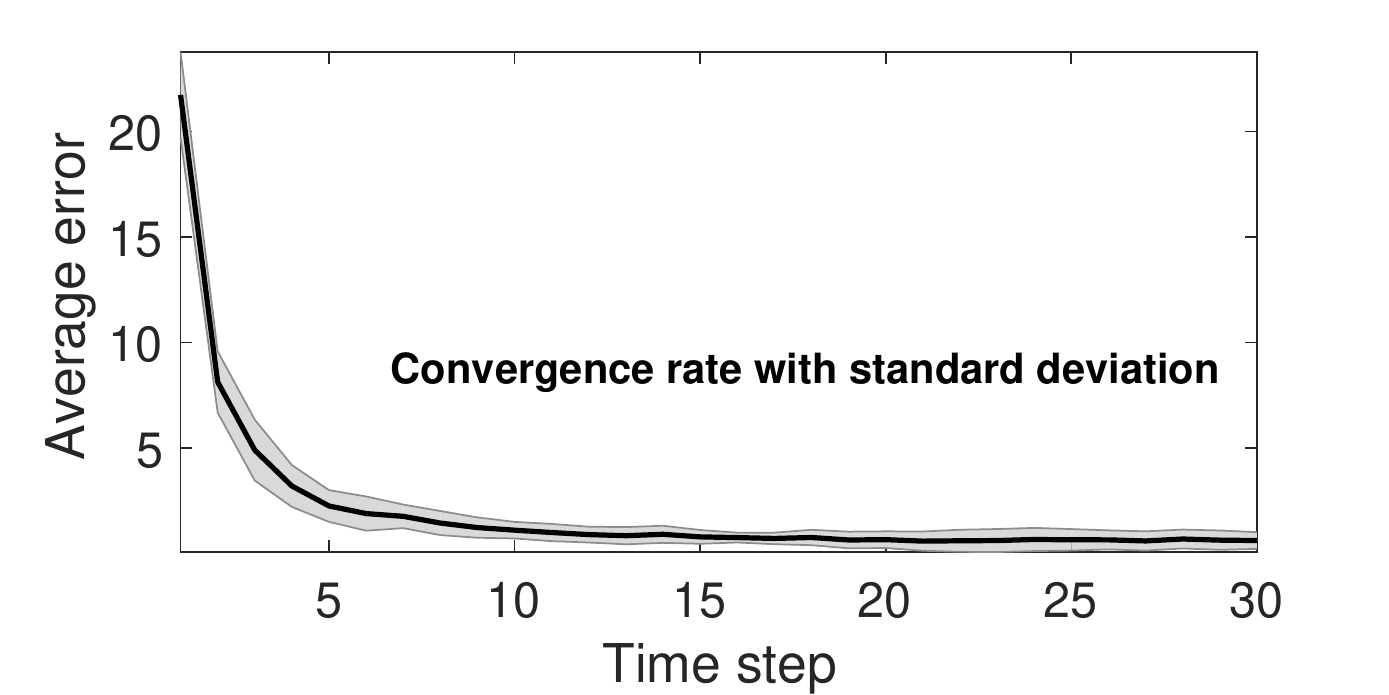}}\\
\subfloat[Comparison with deterministic Distributed Averaging]{\label{static:2}\includegraphics[width=0.45\textwidth]{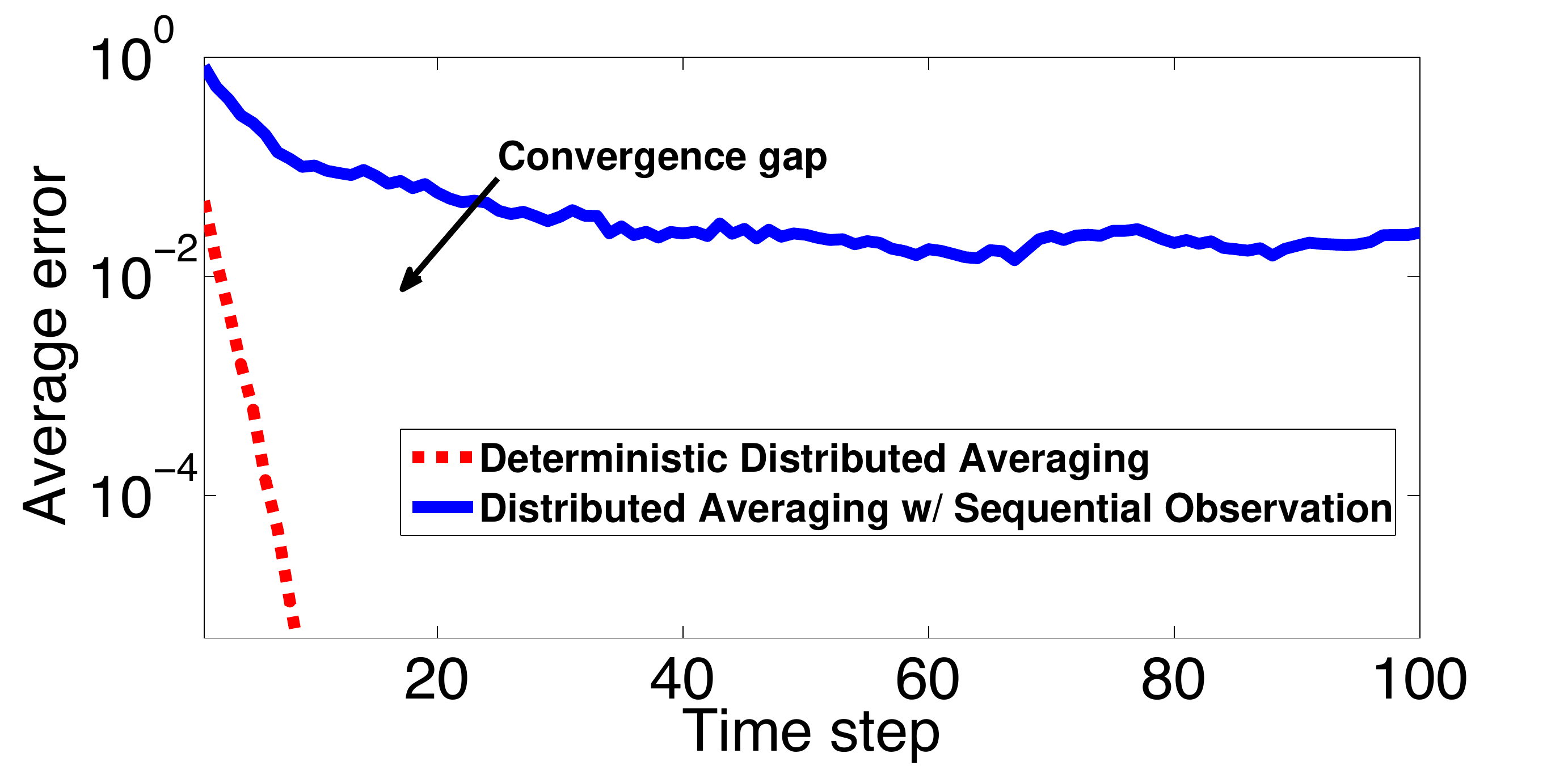}}
\caption{Performance evaluation over  a static undirected graph}
\end{figure}

It can  be readily seen that our algorithm converges nicely with bounded variance. Moreover, comparing its convergence with that of the deterministic distributed averaging algorithm, where true beliefs are revealed to each agent at the initial time step, we see a clear (order-wise) gap between these two scenarios, which is primarily due to the sequential arriving nature of the data in our sequential data arriving setting (notice that in Fig. \ref{static:2} we have changed the y-axis to log scale). This also validates the proved ${O}(1/t)$  convergence rate of the algorithm, which is order-wise slower than the deterministic scenario (which is exponentially fast).

We then investigate the convergence  performance over a time-varying  undirected graph. 
Following \cite{el2016distributed}, 
we generate random probabilistic dynamic-graphs on the basis of a connected \emph{union} graph. In particular, given  a union graph $(\scr{N},\scr{E})$, the agents $i$ and $j$ with  $(i,j)\in\scr{E}$ are connected  with  probability $p$, generating a time-varying $\bbb{N}(t)$. We make sure that the generated graph is connected. \kai{Note that such time-varying graphs satisfy both Assumptions \ref{assum:Joint_Con} and \ref{assum:dyna_graph_value}}. The Metropolis weights are then calculated over this  $\bbb{N}(t)$ at each time $t$. 
As shown in Fig. \ref{dyn:1}, a higher value of $p$ leads to a faster convergence rate as expected. Moreover, a smaller variance is incurred when the graph has less variability over time. In any case, the  polynomial  convergence rate shown in Theorem \ref{thm:convergence} is corroborated.



\begin{figure}[!ht]
\centering
\subfloat{\includegraphics[width=0.45\textwidth]{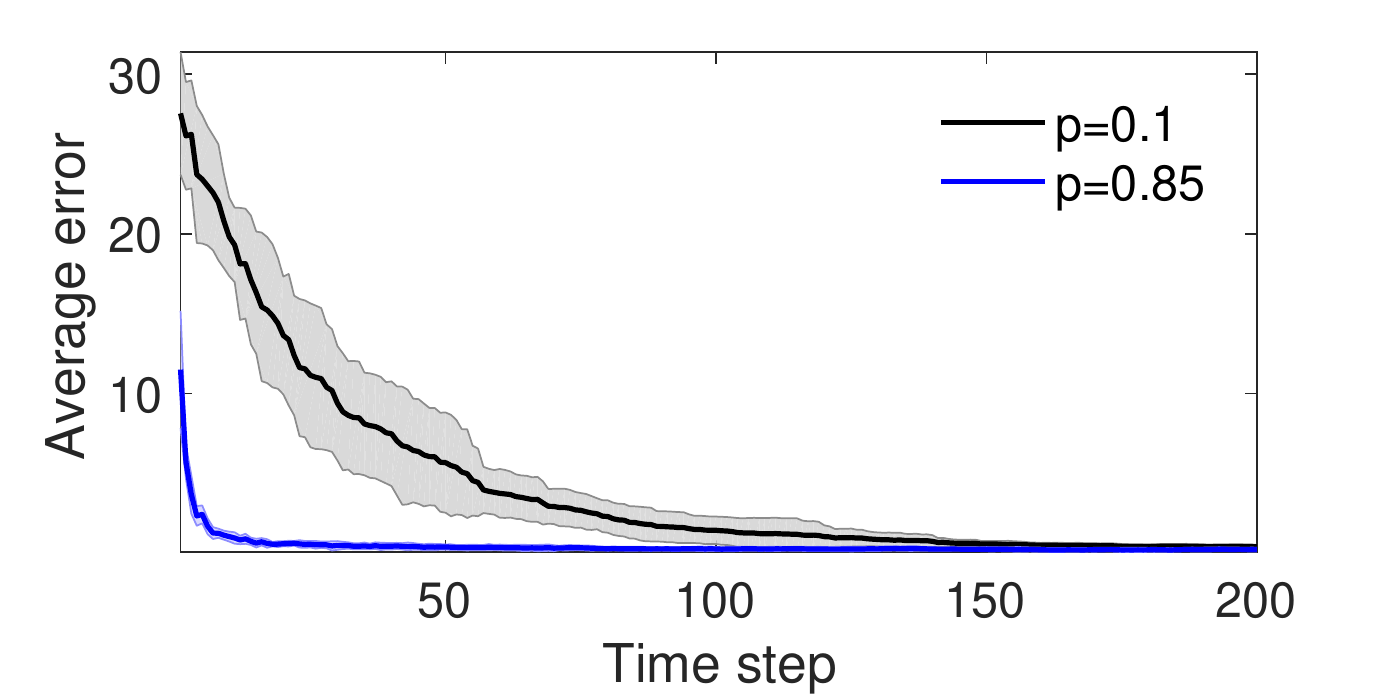}}
\caption{Performance evaluation over  a time-varying undirected graph}\label{dyn:1}
\end{figure}

\subsection{Directed Graphs}
We next  investigate the convergence performance over directed graphs. To generate random directed graphs, 
we first generate undirected RGGs and then randomly delete some of the unidirectional edges between agents. 
With a fixed union graph, a time-varying graph is generated in the same way as in Section \ref{sec:sim_undirected}.  \kai{Thus, the graphs satisfy Assumption \ref{assum:Joint_Con_Strong}.}
We have tested the update \eqref{update_direct} for both static and time-varying graphs. It can be seen in Fig. \ref{fig:directed}   that the network-wide belief averaging is successfully achieved at polynomial rate, which corroborates the theoretical results in Theorem \ref{thm:converge_DG}. Similarly, a higher value of $p$ results in  a faster convergence rate and a smaller variance.

\begin{figure}[!ht]
\centering
\subfloat[Convergence of the average error over a static graph]{\label{fig:directed_1}\includegraphics[width=0.45\textwidth]{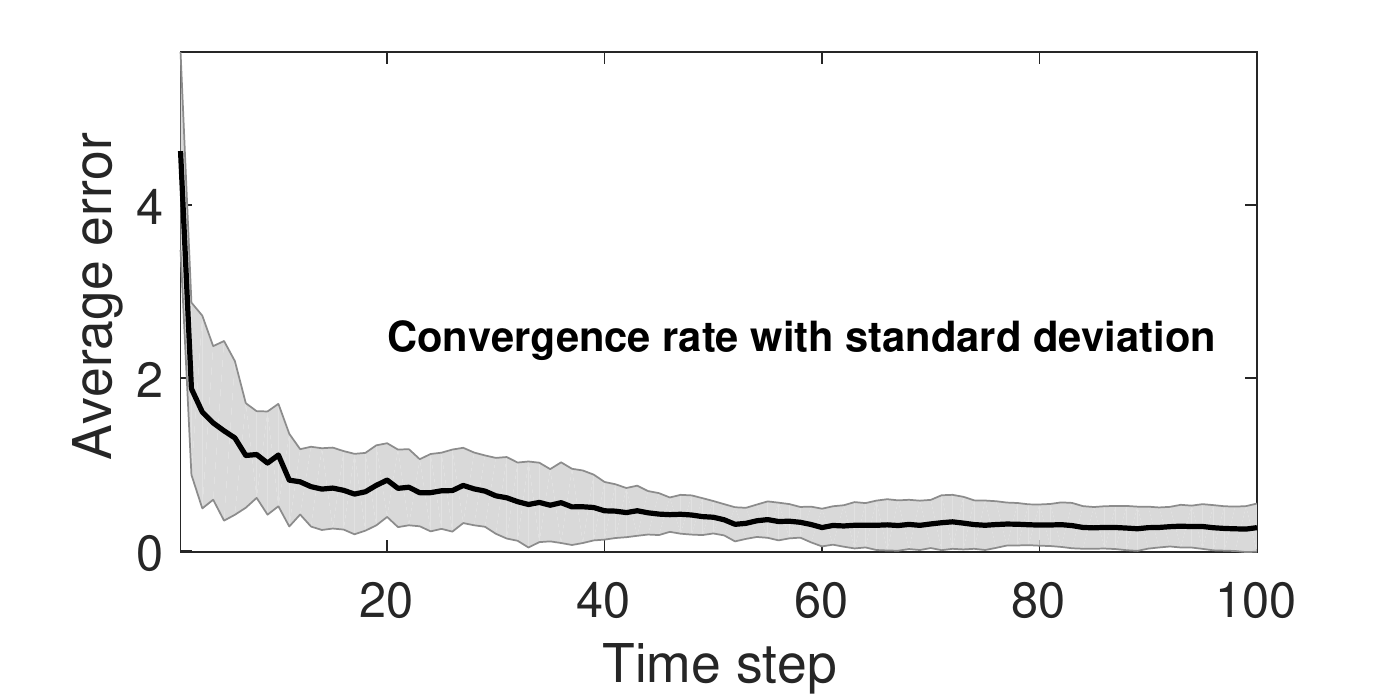}}\\ 
\subfloat[Convergence of the average error over a time-varying graph]{\label{fig:directed_2}\includegraphics[width=0.45\textwidth]{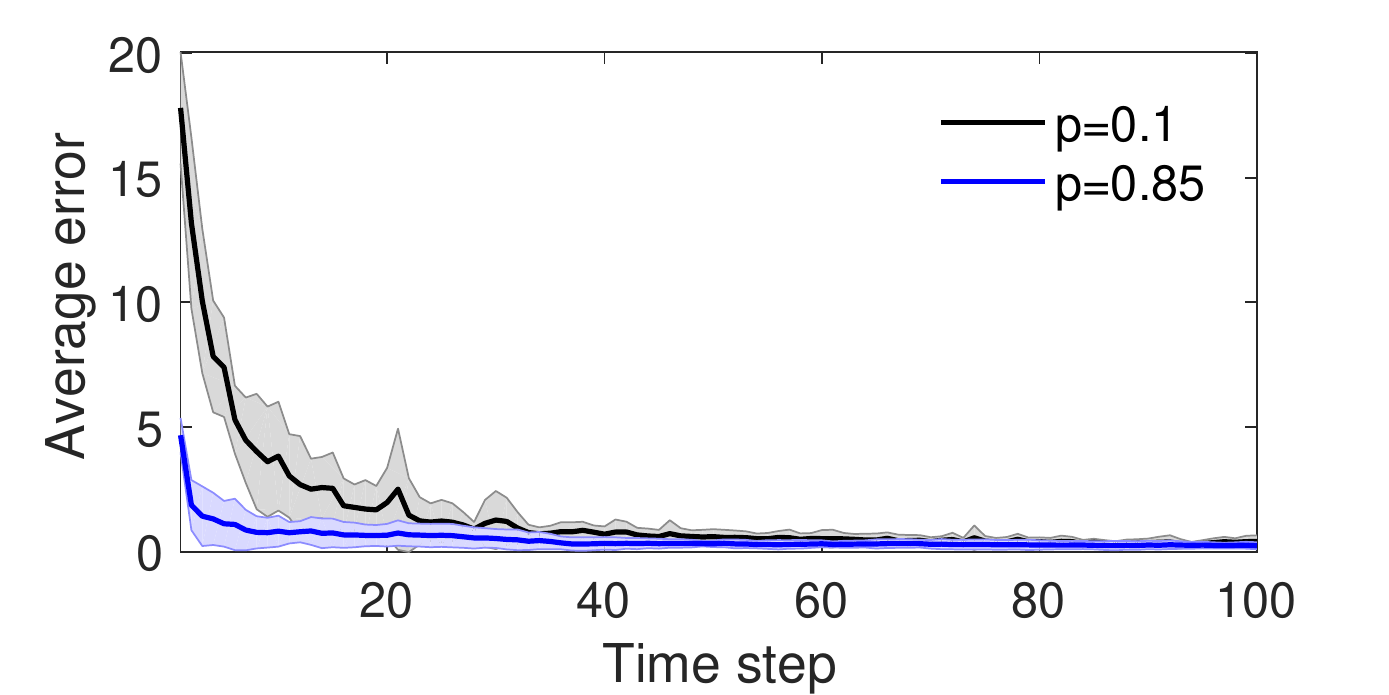}}
\caption{Performance evaluation over  a  directed graph}\label{fig:directed}
\end{figure}

\subsection{Quantization}
In addition, we also study the effects of the two sources of quantization considered in Section \ref{sec:quan}. 
To satisfy conditions A.1) and A.2) in  Assumption \ref{assum:quan_weight}, we adopt a modified Metropolis weight as in \cite{chamie}. Specifically, we let 
\begin{align}
\label{equ:mod_Metropolis_1}
w_{ij}(t) &= \frac{1}{C(1+\max\{n_i(t),n_j(t)\})}~, \;\; (i,j)\in \bbb{N}(t)~,
\end{align}
with $C>1$ 
and $w_{ii}(t)$  selected such that  $\sum_{j=1}^n w_{ij}(t)=1$. Moreover, we choose the communication quantizer $\scr{Q}$ to be the truncation operator, and the precision\footnote{Note that in practice, the precision may be smaller than $0.1$. We choose such a relatively large value just to better illustrate the convergence results.} $\Delta$ to be $0.1$. We first consider the case when the local belief $\bar{x}_t\notin\scr{B}$ for any $i\in\scr{N}$. It is demonstrated  in Fig. \ref{fig:quan_static} that over a static undirected graph, the  average error $e(t)$  indeed converges in around $130$ iterations under the quantization we consider. In contrast to the results without quantization, however, the convergence of the error $e(t)$ is stalled at somewhere above zero (note that we have log scale y-axis in Fig. \ref{fig:quan_static_1}). In fact,  as shown in Fig. \ref{fig:quan_static_2}, the local state $y_i(t)$ converges to the neighborhood of the average belief at a very fast rate. In the middle of convergence, oscillation of the states is observed, which is similar to the  cyclic behavior as reported in \cite{chamie}. Due to the stochastic nature of the sequential  data, the  oscillation may not be exact cyclic. Eventually, the local state values converge to the quantized consensus that deviates from the actual average belief by less than $1$. More examples have been observed to have  similar convergence results as shown in Fig. \ref{fig:quan_static},   which corroborate the convergence results in Proposition \ref{prop:quan_conv_static}.

\begin{figure}[!ht]
\centering
\subfloat[Convergence of the average error]{\label{fig:quan_static_1}\includegraphics[width=0.45\textwidth]{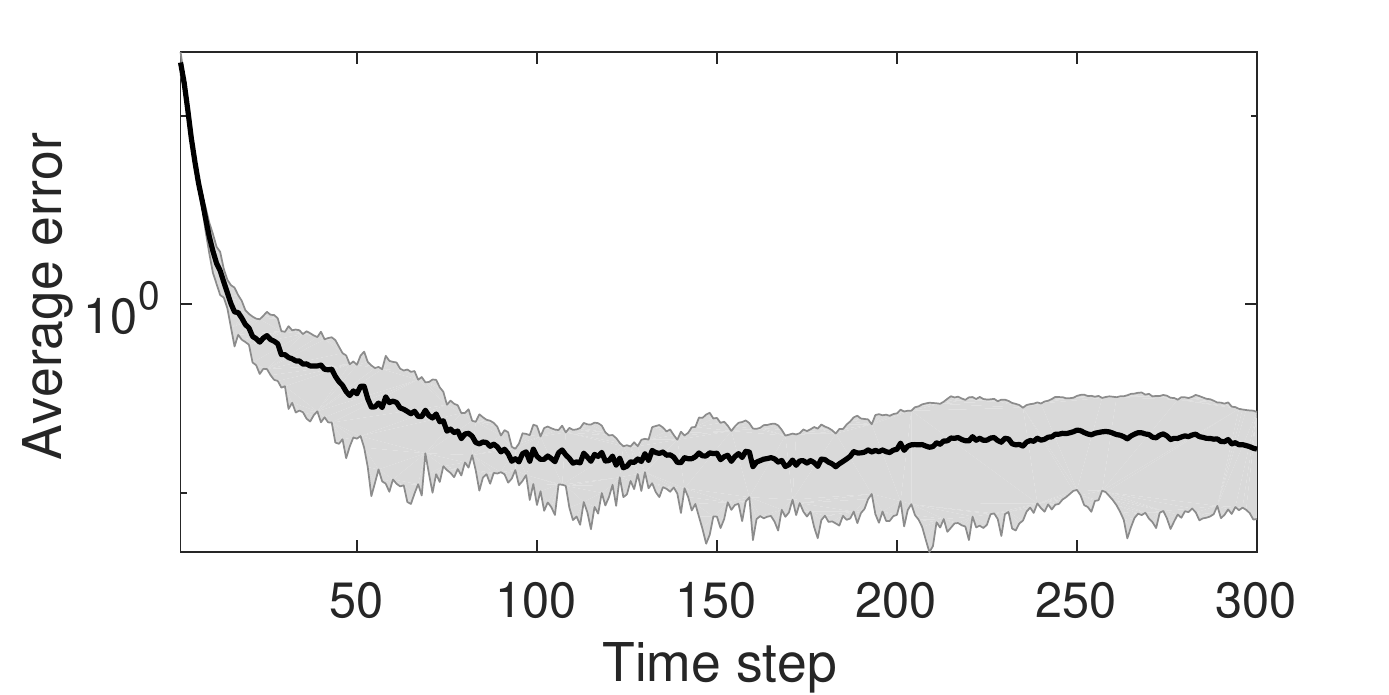}}\\
\subfloat[Convergence of the local state $y_i(t)$]{\label{fig:quan_static_2}\includegraphics[width=0.45\textwidth]{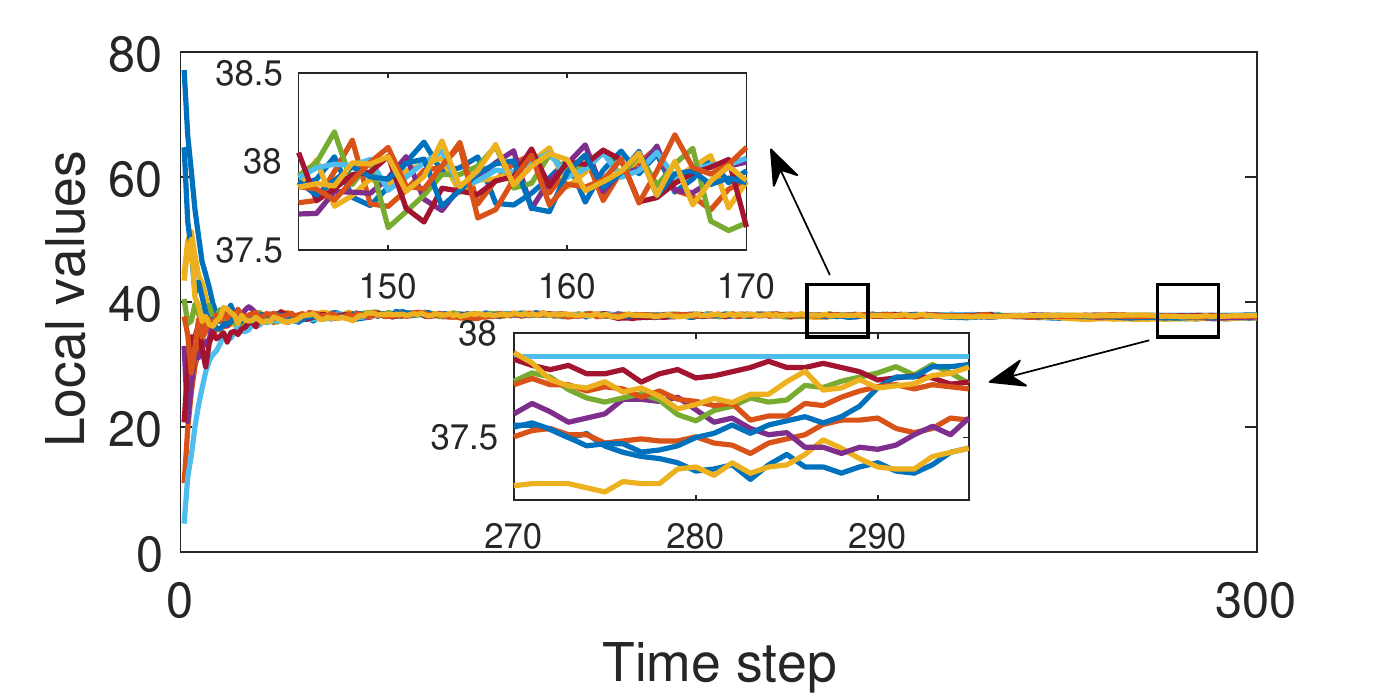}}
\caption{Performance evaluation under  quantization over a static graph. In (b), each curve corresponds to the evolution of  local state $y_i(t)$ at each agent.}\label{fig:quan_static}
\end{figure}

\begin{figure}[!ht]
\centering
\subfloat[Convergence of the average error]{\label{fig:quan_dyna_1}\includegraphics[width=0.45\textwidth]{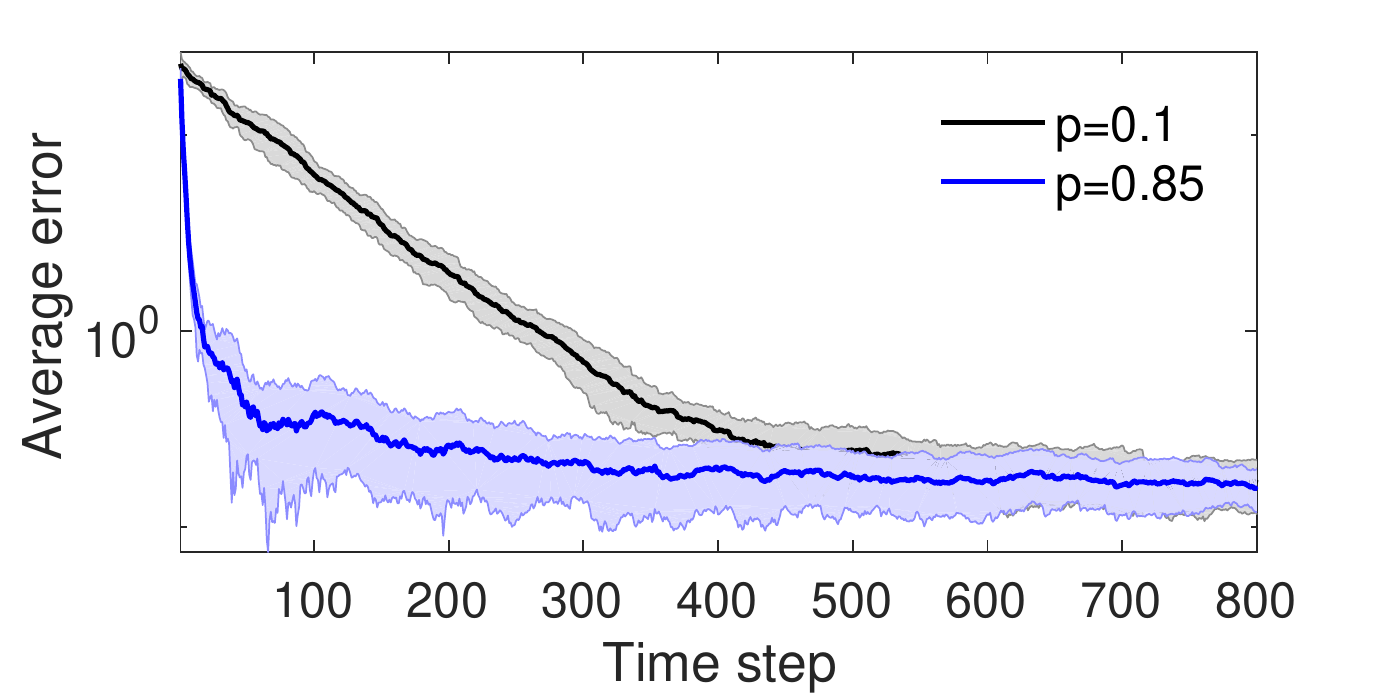}}\\
\subfloat[Convergence of the local state $y_i(t)$ with $p=0.85$]{\label{fig:quan_dyna_2}\includegraphics[width=0.45\textwidth]{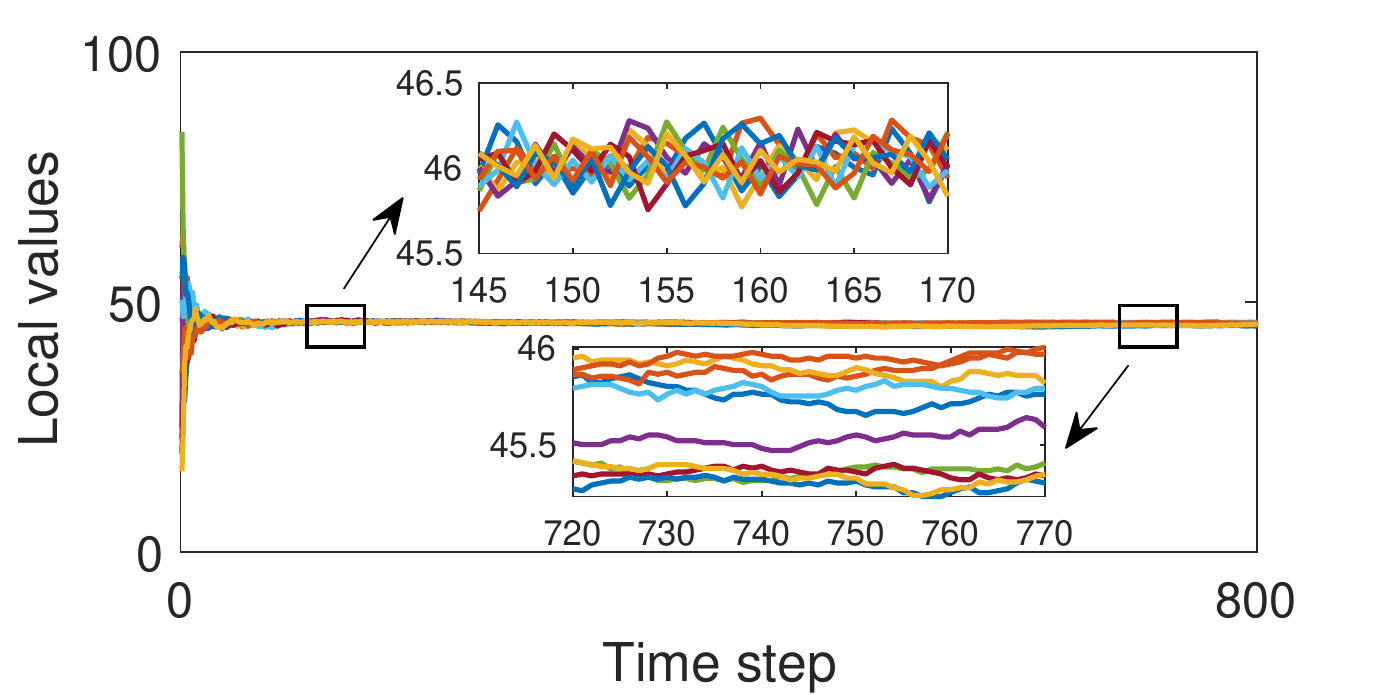}}
\caption{Performance evaluation under  quantization over a time-varying  graph. In (b), each curve corresponds to the evolution of  local state $y_i(t)$ at each agent.}\label{fig:quan_dyna}
\end{figure}

\begin{figure}[!ht]
\centering
\subfloat[Convergence of the average error]{\label{fig:quan_dyna_B_set_1}\includegraphics[width=0.45\textwidth]{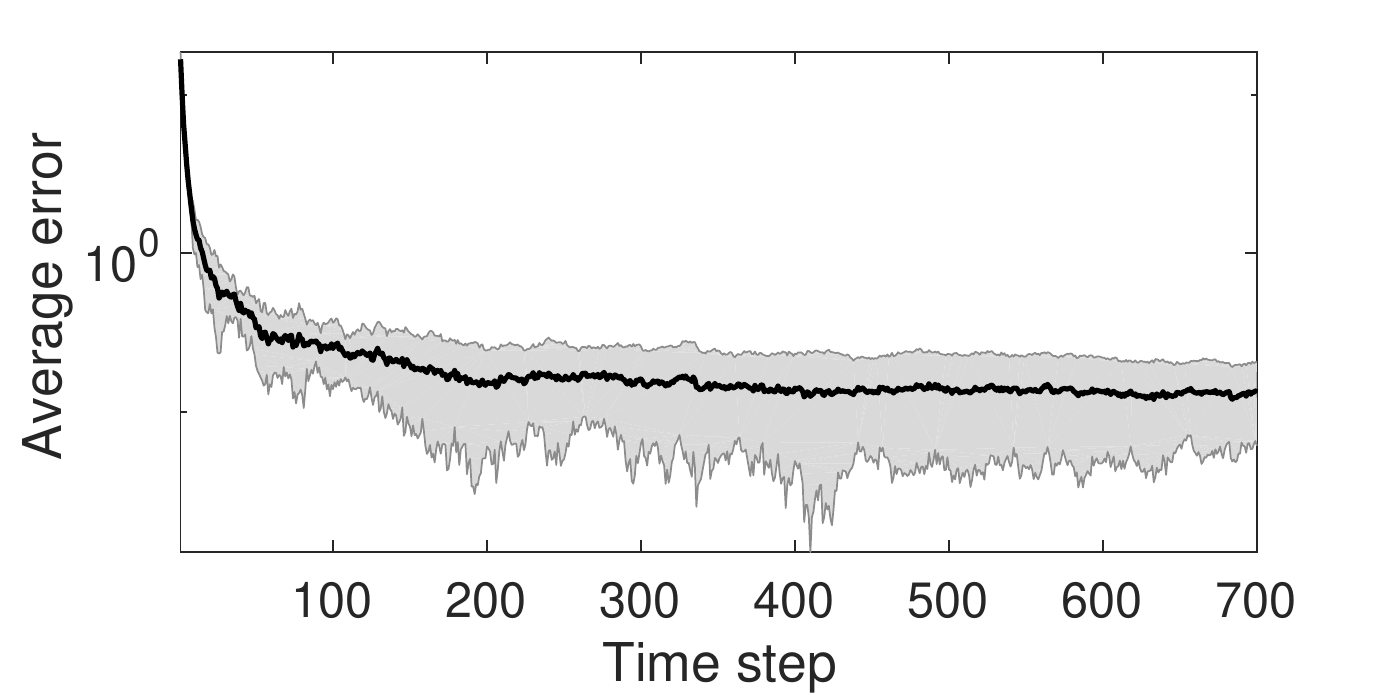}}\\
\subfloat[Convergence of the local state $y_i(t)$]{\label{fig:quan_dyna_B_set_2}\includegraphics[width=0.45\textwidth]{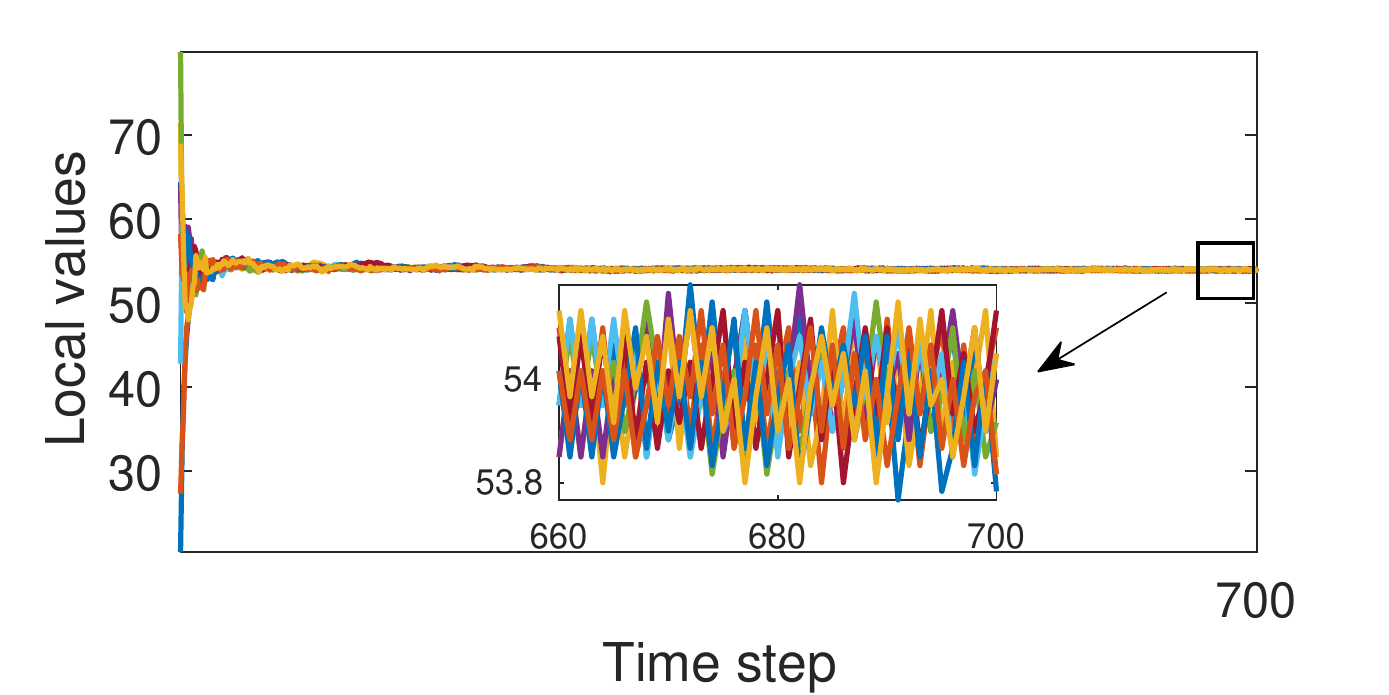}}
\caption{Performance evaluation under  quantization when $\bar{x}_i\in\scr{B}$ for all $i\in\scr{N}$. In (b), each curve corresponds to the evolution of  local state $y_i(t)$ at each agent.}\label{fig:quan_dyna_B_set}
\end{figure}

Likewise, as shown in Fig. \ref{fig:quan_dyna_1}, over a time-varying graph, the average error $e(t)$ converges (in around $300$ and $600$ iterations for $p=0.85$ and $p=0.1$, respectively), at a relatively slower rate than the case with a static graph. 
Moreover, the error $e(t)$ still 
fails to converge to exactly zero due to the  quantization effects. Furthermore, the convergence behavior of the local states $y_i(t)$ in Fig. \ref{fig:quan_dyna_2} resembles what is shown in Fig. \ref{fig:quan_static_2}, while it takes longer time to converge to  quantized consensus. 
Additionally, we are also interested in the convergence performance when some  local beliefs $\bar{x}_i$ are inside the belief set $\scr{B}$. We thus specifically  round the random beliefs $\bar{x}_i$ to the set $\scr{B}$ by finding  the value in $\scr{B}$ that is closest to $\bar{x}_i$ in magnitude. Fig. \ref{fig:quan_dyna_B_set_1} illustrates that the local states also fail to reach exact consensus to the average belief. Interestingly, the states here do not achieve the quantized consensus as in Fig. \ref{fig:quan_static_2} and Fig.\ref{fig:quan_dyna_2}, while keep oscillating (though not exact cycling) around the neighborhood of the consensual belief. This somehow reflects the difficulty we encountered in theoretical analysis, that the difference of the running average with limited precision will randomly take values from  $\Delta$, $-\Delta$, or $0$ (see Remark \ref{remark:B_set}). The random error $\Delta \tilde{z}_i(t)$ does not accumulate over time and 
the size of the neighborhood is bounded to be within $1$.

{\color{black}
\subsection{Convergence Speed}\label{sec:conv_rate_quan}
We also numerically investigate how the convergence speed is influenced  by the quantization effects. As shown in Fig. \ref{fig:quan_rate_comp}, we compare the convergence of the average errors under various levels of quantization. Note that $\Delta=0$ represents the case where only communication quantization exists and the division operation leads to no precision errors. As expected, a higher  level of quantization leads to  a slower  convergence  rate and a larger steady-state error. 
Surprisingly,  however, the convergence rate is insensitive to  either sources of quantizations. This implies that   $O(1/t)$ seems to be  an inherent convergence rate in the distributed averaging problem using  sequential data. Thus, we conjecture  that the rate to reach consensus under quantization is still in the order of $O(1/t)$, whose proof is left for future work.

\begin{figure}[!ht]
\centering
\subfloat{\includegraphics[width=0.45\textwidth]{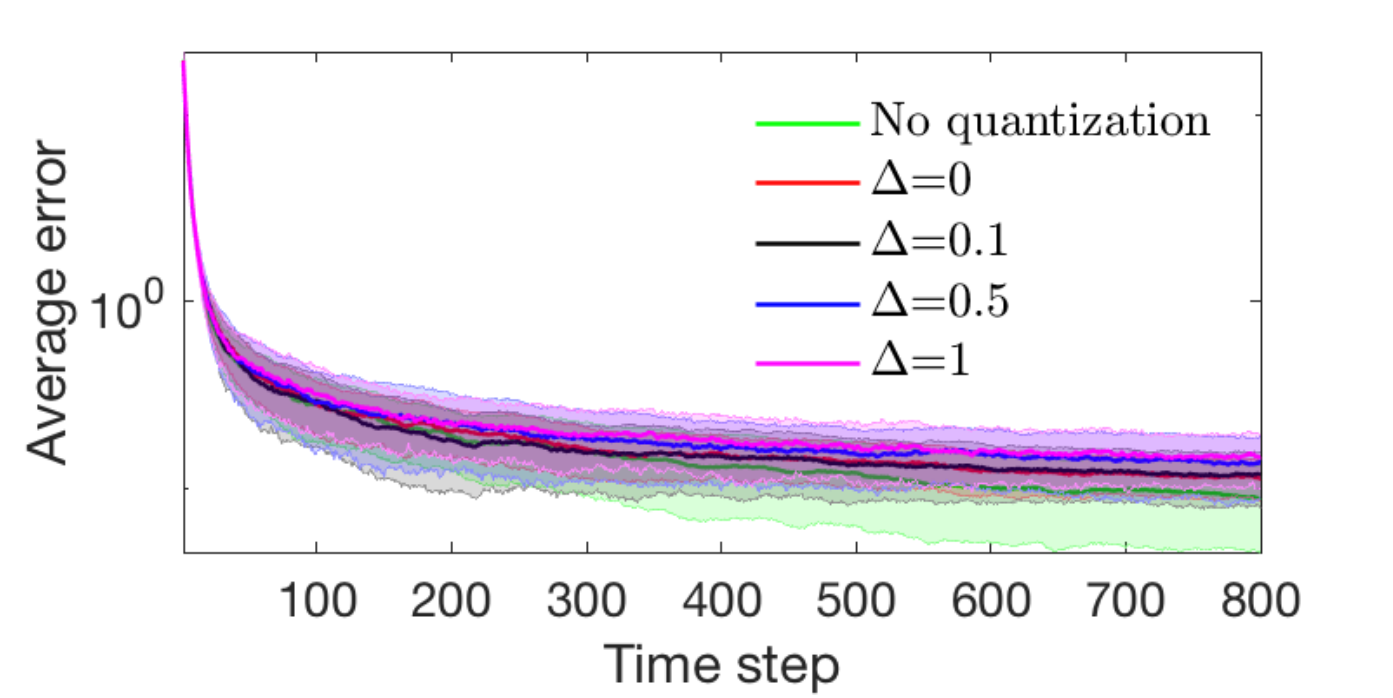}}
\caption{Performance evaluation under various levels of quantizations.}\label{fig:quan_rate_comp}
\end{figure} 

}

\section{Conclusion}\label{sec:conclude}

In this paper, we have studied  distributed learning of average belief  over networks with sequential observations, where in contrast to
the conventional distributed averaging problem setting, each
agent's local belief is not available immediately and can only
be learned through its own observations. 
Two distributed algorithms
have been introduced for solving the problem over time-varying undirected and directed graphs, and their polynomial convergence rates have been established. 
We have also modified the algorithm for the practical case in which both quantized communication and limited precision of division operation occur. 
Numerical results have been provided to corroborate our theoretical findings.

For future work, we plan to  investigate other important aspects of the proposed scheme for distributed learning of average belief with sequential data, e.g., the case under malicious data attack or with privacy requirement among agents. It is also interesting to connect the proposed scheme with other distributed and multi-agent  learning algorithms \cite{shamir2014fundamental,rahimian2016distributed,zhang2017projection,zhang2018fully,zhang18cdc}.

\begin{ack}                               
The work of K. Zhang and T. Ba\c{s}ar
was supported in part by  Office of Naval Research (ONR) MURI Grant N00014-16-1-2710, and in part by US Army Research Office (ARO) Grant W911NF-16-1-0485.  The work of M. Liu is partially supported by the NSF under grants ECCS 1446521, CNS-1646019, and CNS-1739517. The authors also thank Zhuoran Yang from Princeton University for the helpful discussion that improves the final version of the paper.
\end{ack}

\bibliographystyle{unsrt}
\bibliography{ji}

\end{document}